\documentclass[usenatbib,useAMS,usedcolumn]{mn2e}

\usepackage{epsfig}
\usepackage{times}

\newcommand{\arcm}{\hbox{$^\prime$}}

\newcommand{\degree}{\hbox{$^\circ$}}

\newcommand{\chandra}{\emph{Chandra}}
\newcommand{\xmm}{\emph{XMM-Newton}}
\newcommand{\xmms}{\emph{XMM}}

\newcommand{\arcs}{\mbox{\arcm\arcm}}
\newcommand{\Lx}{\ensuremath{L_{\mathrm{X}}}}
\newcommand{\Tx}{\ensuremath{T_{\mathrm{X}}}}
\newcommand{\Zsol}{\ensuremath{Z_{\odot}}}
\newcommand{\Lsol}{\ensuremath{L_{\odot}}}
\newcommand{\Msol}{\ensuremath{M_{\odot}}}
\newcommand{\LB}{\ensuremath{L_{\mathrm{B}}}}
\newcommand{\LBsol}{\ensuremath{L_{B\odot}}}

\newcommand{\Bfit}{\ensuremath{\beta_{fit}}}
\newcommand{\Bfittwo}{\ensuremath{\beta_{fit,2}}}

\newcommand{\sigT}{\ensuremath{\mbox{$\sigma$:\Tx}}}

\newcommand{\NH}{\ensuremath{N_{\mathrm{H}}}}

\newcommand{\s}{\ensuremath{\mbox{~s}}}
\newcommand{\ps}{\ensuremath{\s^{-1}}}
\newcommand{\cm}{\ensuremath{\mbox{~cm}}}
\newcommand{\pcmsq}{\ensuremath{\cm^{-2}}}
\newcommand{\km}{\ensuremath{\mbox{~km}}}
\newcommand{\Mpc}{\ensuremath{\mbox{~Mpc}}}
\newcommand{\pMpc}{\ensuremath{\Mpc^{-1}}}
\newcommand{\kmpspMpc}{\ensuremath{\km \ps \pMpc\,}}
\newcommand{\erg}{\ensuremath{\mbox{~erg}}}
\newcommand{\ergps}{\ensuremath{\erg \ps}}
\newcommand{\ergpspcmsq}{\ensuremath{\erg \ps \pcmsq}}
\newcommand{\kmps}{\ensuremath{\km \ps}}

\newcommand{\Ho}{\ensuremath{H_\mathrm{0}}}

\newcommand{\ML}{\ensuremath{\mbox{\Msol/\LBsol}}}

\newcommand{\Dtf}{\ensuremath{D_{\mathrm{25}}}}

\newcommand{\xmmu}{\mbox{XMMU J001526.4+171723}}
\newcommand{\gtsim}{\,\rlap{\raise 0.5ex\hbox{$>$}}{\lower 1.0ex\hbox{$\sim$}}\,} 
\begin{document}

\title[ 
Dark haloes of isolated early-type galaxies 
] 
{ 
  The dark haloes of early-type galaxies in low-density environments:
  XMM-Newton and Chandra observations of NGC 57, NGC 7796 and IC
  1531\footnotemark[1]
}

\author[
E. O'Sullivan, A.~J.~R. Sanderson and T.~J. Ponman
]
{
Ewan O'Sullivan\footnotemark[2]$^{1}$, Alastair~J.~R. Sanderson$^{2}$ and Trevor~J. Ponman$^{2}$\\
$^{1}$ Harvard-Smithsonian Center for Astrophysics, 60 Garden Street, Cambridge, MA 02138 \\ 
$^{2}$ School of Physics and Astronomy, University of Birmingham, Edgbaston, B15 2TT, UK
}

\date{Accepted 2007 July 11. Received 2007 July 10; in original form 2007 March 30}

\pagerange{\pageref{firstpage}--\pageref{lastpage}} \pubyear{2007}

\maketitle

\label{firstpage}

\begin{abstract} 
%200 words, no refs
  We present analysis of \chandra\ and \xmm\ observations of three
  early-type galaxies, NGC~57, NGC~7796 and IC~1531. All three are found in
  very low density environments, and appear to have no neighbours of
  comparable size. NGC~57 has a halo of kT$\sim$0.9~keV, solar metallicity
  gas, while NGC~7796 and IC~1531 both have $\sim$0.55~keV, 0.5-0.6~\Zsol
  haloes. IC~1531 has a relatively compact halo, and we consider it likely
  that gas has been removed from the system by the effects of AGN heating. For
  NGC~57 and NGC~7796 we estimate mass, entropy and cooling time profiles
  and find that NGC~57 has a fairly massive dark halo with a mass-to-light
  ratio of 44.7$^{+4.0}_{-8.5}$~\ML\ (1$\sigma$ uncertainties) at
  4.75$r_e$. This is very similar to the mass-to-light ratio found for
  NGC~4555 and confirms that isolated ellipticals can possess sizable
  dark matter haloes.  We find a significantly lower mass-to-light ratio
  for NGC~7796, 10.6$^{+2.5}_{-2.3}$ \ML\ at 5$r_e$, and discuss the
  possibility that NGC~7796 hosts a galactic wind, causing us to
  underestimate its mass.
\end{abstract}


\footnotetext[1]{Based on observations obtained with XMM-Newton, an ESA
science mission with instruments and contributions directly funded by ESA
Member States and NASA.}
\footnotetext[2]{E-mail: eosullivan@head.cfa.harvard.edu}

\section{Introduction}
\label{sec:intro}
While the majority of early-type galaxies are found in groups and clusters
\citep{Tully87,MelnickSargent77,Dressler80}, there is reason to believe
that these environments may not be best choice for studies of intrinsic
galaxy properties. The low velocities and high densities of galaxy groups
are conducive to galaxy merging, and simulations suggest that gravitational
encounters between galaxies can lead to the loss of dark matter (DM) to the
surrounding group or cluster halo \citep{Barnes89}.  X-ray observations
have shown that the potential wells of groups and clusters are commonly
filled with haloes of high temperature gas \citep{Kelloggetal75}. Elliptical
galaxies are also known to possess such gaseous haloes
\citep{Formanjonestucker85,Trinchierietal86}. As well as the inevitable
confusion caused by attempting to observe a galaxy halo through surrounding
cluster gas, ram-pressure and viscous stripping may remove gas from
individual galaxies as they move through the intra-cluster medium
\citep[ICM, e.g.][]{Acremanetal03}.  Models of massive galaxies in the
cores of larger systems suggest that their haloes may be enhanced by the
inflow of gas from the surrounding intracluster medium
\citep{MathewsBrighenti03}. However, the dominant elliptical galaxies of
clusters and groups are known to be the most X-ray luminous examples of
their type \citep{Helsdonetal01,OFP01cat}, and so are often chosen for the
most detailed X-ray analysis. While very high quality can be obtained from
such observations, they can only be at best a doubtful source of
information on the inherent X-ray properties of early-type galaxies.

One motivation for examining the gas haloes of early-type galaxies is that,
under the assumption that they are in hydrostatic equilibrium, it is
possible to use them to infer the total mass profile of the galaxy. This
feature has become particularly important recently, owing to reports that
some ellipticals may contain little or no dark matter
\citep{Romanowskyetal03}, based on modelling of the kinematics of their
planetary nebula population. These have provoked discussion of the
potential biases affecting different dynamical mass estimation techniques
\citep[e.g.][]{Dekeletal05,MamonLokas05a,Douglasetal07}.  Dynamical mass
measurements at large radii rest on modelling of the velocity field of
planetary nebulae or globular clusters, which may be affected by a number of
factors, most notably orbital anisotropy, flattening of the galaxy
along the line of sight, and the assumed link between planetary nebulae and
the old stellar population, which is uncertain.

Given the uncertainties associated with kinematical modelling, mass
profiles based on X-ray observations can potentially provide a much needed
independent measurement of DM content. However, as described above, group
and cluster environments may affect the properties of member galaxies and
introduce confusion to observations. Isolated ellipticals are therefore
preferable for this type of study. Mass estimation in the inner regions of
ellipticals may also be biased by pressure fluctuations induced by energy
injection from an active nucleus. Analysis of \chandra\ observations of a
number of ellipticals shows that their optical and X-ray morphologies are
uncorrelated within 1.2 optical effective radii \citep{DiehlStatler07}. As
the mass within this radius should be completely dominated by the stellar
contribution, the lack of correlation suggests that the gas is not in
hydrostatic equilibrium, with AGN activity the most likely culprit. For low
mass or dynamically young ellipticals there is also the possibility that
the X-ray emitting gas may take the form of a wind rather than a
hydrostatic halo.  This has been demonstrated in detail for NGC~3379, where
X-ray and dynamical mass estimates disagree strongly
\citep{PellegriniCiotti06}. Modelling as an outflowing wind with properties
determined by the rate of supernova heating as well as the mass profile
produces a closer agreement . We therefore need to exercise caution when
choosing ellipticals for study and when interpreting the results, to
account for the potential errors introduced by gas motions and AGN heating,
particularly in lower mass systems.

In a previous paper \citep[][hereafter referred to as
OP04]{OSullivanPonman04b_special}, we discussed a \chandra\ observation of
NGC~4555, an isolated elliptical galaxy with an extended X-ray halo, and
showed that it possessed a dark matter halo. In this paper we examine \xmm\
and \chandra\ observations of three other early-type galaxies in
low-density environments, NGC~57, NGC~7796 and IC~1531. All three galaxies
were selected using the same strict isolation criteria used to identify NGC~4555 (discussed in \textsection~\ref{sec:environs}), and NGC~57 has been
identified as isolated by other surveys \citep{Smithetal04}.  IC~1531 hosts
a fairly bright extended radio source \citep{Ekersetal89} apparently
associated with the active nucleus of the galaxy. NGC~57 and NGC~7796 have
no known radio sources associated with them. None of the galaxies have been
extensively studied in the X-ray, and optical data for NGC~57 and IC~1531
are sparse. \citet{Bertinetal94} found NGC~7796 to be a boxy elliptical
with a small rotational velocity and flat velocity dispersion profile. The
galaxy does possess a counter-rotating core inside 4\arcs, consisting of
old, $\alpha$-element enhanced stars \citep{Miloneetal07}, perhaps
indicating a past interaction or merger. Despite this, NGC~7796 lies on the
Fundamental Plane \citep{Redaetal05}, suggesting that the stellar component
is relaxed, and has a luminosity weighted age of 11.8$\pm$1.9~Gyr
\citep{Thomasetal05}.

Some basic details of the three galaxies are given in
Table~\ref{tab:basic}. For NGC~57 and IC~1531 we take distances and
luminosities from \citet{OFP01cat}. For NGC~7796 we adopt the surface
brightness fluctuation distance from \citet{Tonryetal01}, corrected to
match the Cepheid zero point of \citet{Freedmanetal01} as described in
\citet{Jensenetal03}. We correct the \citet{OFP01cat} optical luminosity to
match this distance. Throughout this paper we assume \Ho=75 \kmpspMpc,
consistent with the WMAP 3-year mean value of 74$\pm$3 \kmpspMpc\
\citep{Spergeletal07}, and normalize optical $B$-Band luminosities to the
$B$-band luminosity of the sun, \LBsol=5.2$\times$10$^{32}$\ergps.
Abundances are measured relative to the solar abundance ratios of
\citet{GrevesseSauval98}, which differ from the older ratios of
\citet{AndersGrevesse89} primarily in that they use an abundance of Fe a
factor of $\sim$1.4 lower. 
%We assume $\Omega_M$=0.27 and $\Omega_\Lambda$=0.73, ...

\begin{table}
\begin{center}
\begin{tabular}{lccc}
Galaxy & NGC~57 & IC~1531 & NGC~7796 \\
\hline
R.A. (J2000) & 00$^h$15$^m$30$^s$.9 & 00$^h$09$^m$35$^s$.5 & 23$^h$58$^m$59$^s$.8 \\
Dec. (J2000) & +17\degree19\arcm42\arcs\ & -32\degree16\arcm37\arcs\ & -55\degree27\arcm30\arcs\ \\
Redshift & 0.0181 & 0.0256 & 0.0110\\
Distance (Mpc) & 72.53  & 109.81 & 46.34 \\
1\arcmin\ = (kpc)& 21.10  & 31.94  & 13.48 \\
\Dtf\ radius (kpc)& 25.31  & 27.14  & 15.09 \\
\NH\ (cm$^{-2}$)& 4.01$\times$10$^{20}$ & 1.35$\times$10$^{20}$ & 2.25$\times$10$^{20}$\\
Log \LB\ (\LBsol) & 10.61 & 10.87 & 10.62 \\
\xmms\ ObsID & 0202190201 & 0202190301 & - \\
\chandra\ ObsID & - & 5783 & 7061,7041 \\
\end{tabular}
\end{center}
\caption{\label{tab:basic} Location and scale of the three galaxies, and
  ObIDs for the X-ray exposures used in each case.}
\end{table}

\section{Observations and Data Analysis}
NGC~57 and IC~1531 were observed by \xmm\ during orbits 737 (2003 December
17-18, ObsID 0202190201) and 814 (2004 May 20, ObsID 0202190301), for
$\sim$28.6 ks and $\sim$24.6 ks respectively. A detailed summary of the
\xmm\ mission and instrumentation can be found in \citet{Jansenetal01}, and
references therein. In both cases, the EPIC MOS and PN instruments were
operated in full frame and extended full frame mode, with the medium
filter. The raw data from the EPIC instruments were processed with the
\xmm\ Science Analysis System (\textsc{sas} v. 7.0), using the
\texttt{epchain} and \texttt{emchain} tasks. After filtering for bad pixels
and columns, X--ray events corresponding to patterns 0-12 for the two MOS
cameras and patterns 0-4 for the PN camera were accepted. Both observations
suffered from mild background flaring, and times when the total count rate
deviated from the mean by more than 3$\sigma$ were therefore excluded. The
effective exposure times for the MOS 1, MOS 2 and PN cameras were 21.2,
21.1 and 18.7 ks respectively for NGC~57 and 16.2, 16.3 and 11.7 ks for
IC~1531.

Images and spectra were extracted, responses generated, and point sources
removed as described in \citet{OSullivanetal05}. NGC~57 and IC~1531 cover
small enough areas of the field of view that local background spectra can
also be used, but we also extracted background images and spectra as
described in \citet{OSullivanetal06}, using the ``double-subtraction''
method \citep{Arnaudetal02,Prattetal01}. Comparison showed little
difference between the alternative background spectra. The local background
spectra were used in spectral fitting, and produced generally acceptable
results.

IC~1531 and NGC~7796 were observed by \chandra\ on 2005 August 21 for
$\sim$40~ksec (IC~1531) and on 2006 August 28 and 2006 September 03 for
$\sim$54 and $\sim$20~ksec (NGC~7796), with ACIS operating in Very Faint
mode (ObsIDs 5783, 7061 and 7401). A summary of the \chandra\ mission and
instrumentation can be found in \citet{Weisskopfetal02}. The S3 chip was at
the telescope focus. Reduction and analysis were performed using methods
similar to those described in OP04. The level 1
events file was reprocessed using \textsc{ciao} v3.3 and \textsc{caldb}
v3.2.1 \citep{Fruscioneetal06}. Bad pixels and events with \textsc{ASCA}
grades 1, 5 and 7 were removed. The data were corrected to the appropriate
gain map, the standard time-dependent gain correction was made, and a
background light curve was produced. A background flare occurred toward the
end of the IC~1531 exposure and all periods where the count rate deviated
from the mean by $>3\sigma$ were excluded, leaving a clean exposure time of
30.7~ksec. In the case of NGC~7796, only minor variations were found in the
background, and the useful exposure times were 52.2~ksec and 17.6~ksec. As
the two exposures were taken only a few days apart and have identical
instrumental setup, we merged the events lists before further analysis.
Subsequent comparison of results from the merged data and individual
exposures shows no significant differences.  In all cases, We chose to use
data from the S3 chip only, as the sources are relatively compact and do
not produce any useful flux on the neighbouring chips.

%Figure moved up from next section to force latex to behave itself
\begin{figure*}
\centerline{
\psfig{file=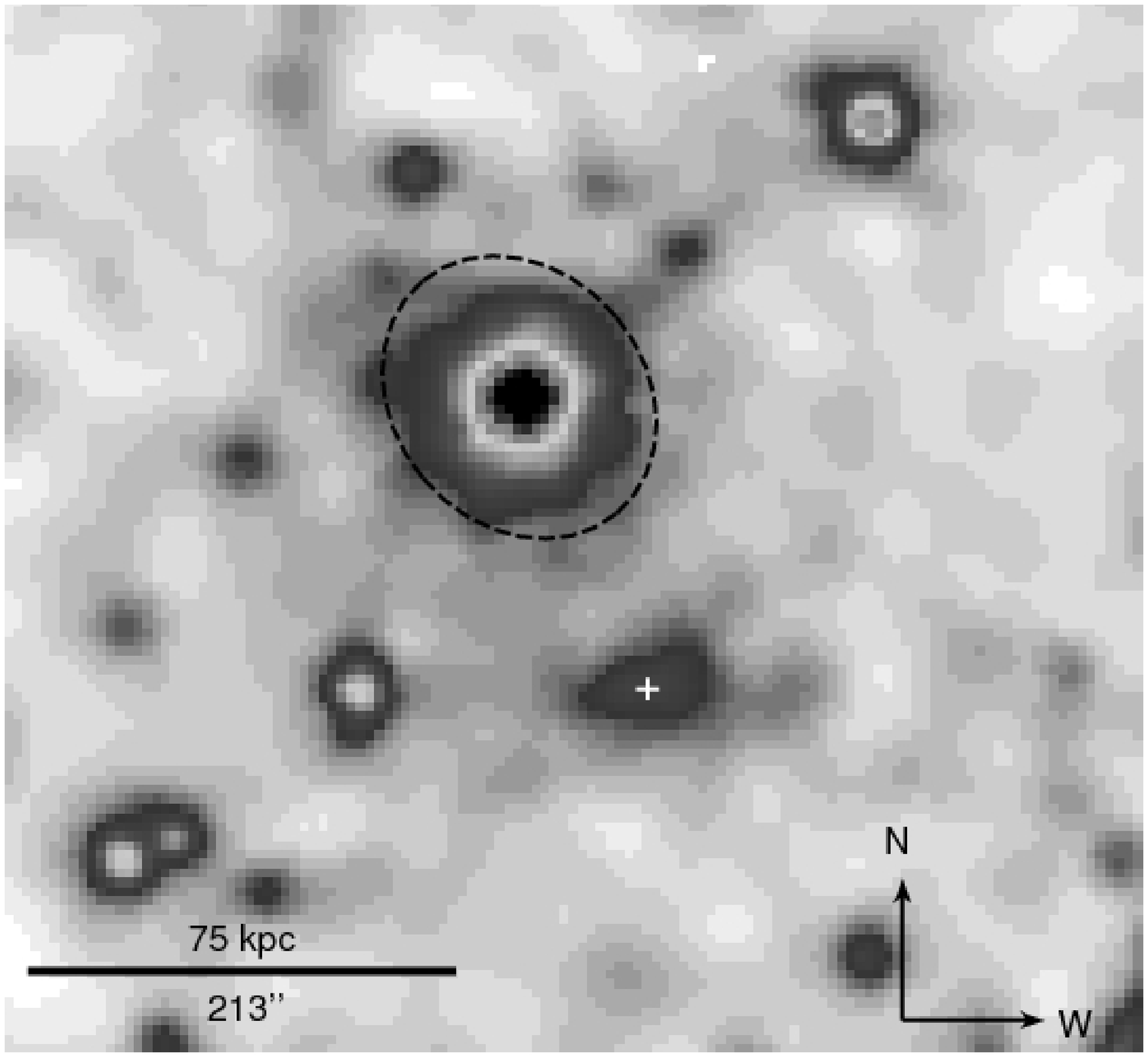,width=6cm}
\psfig{file=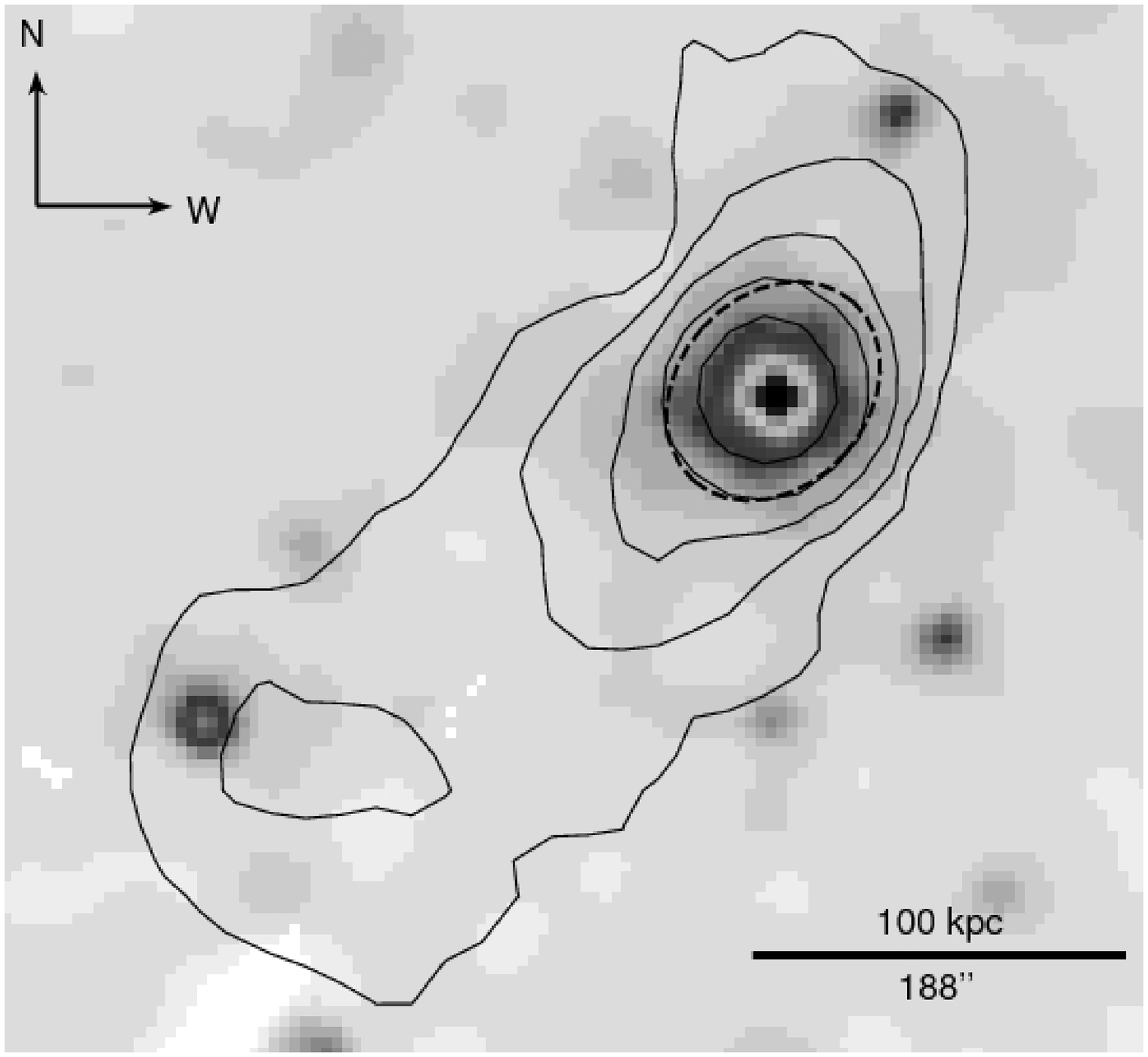,width=6cm}
\frame{\psfig{file=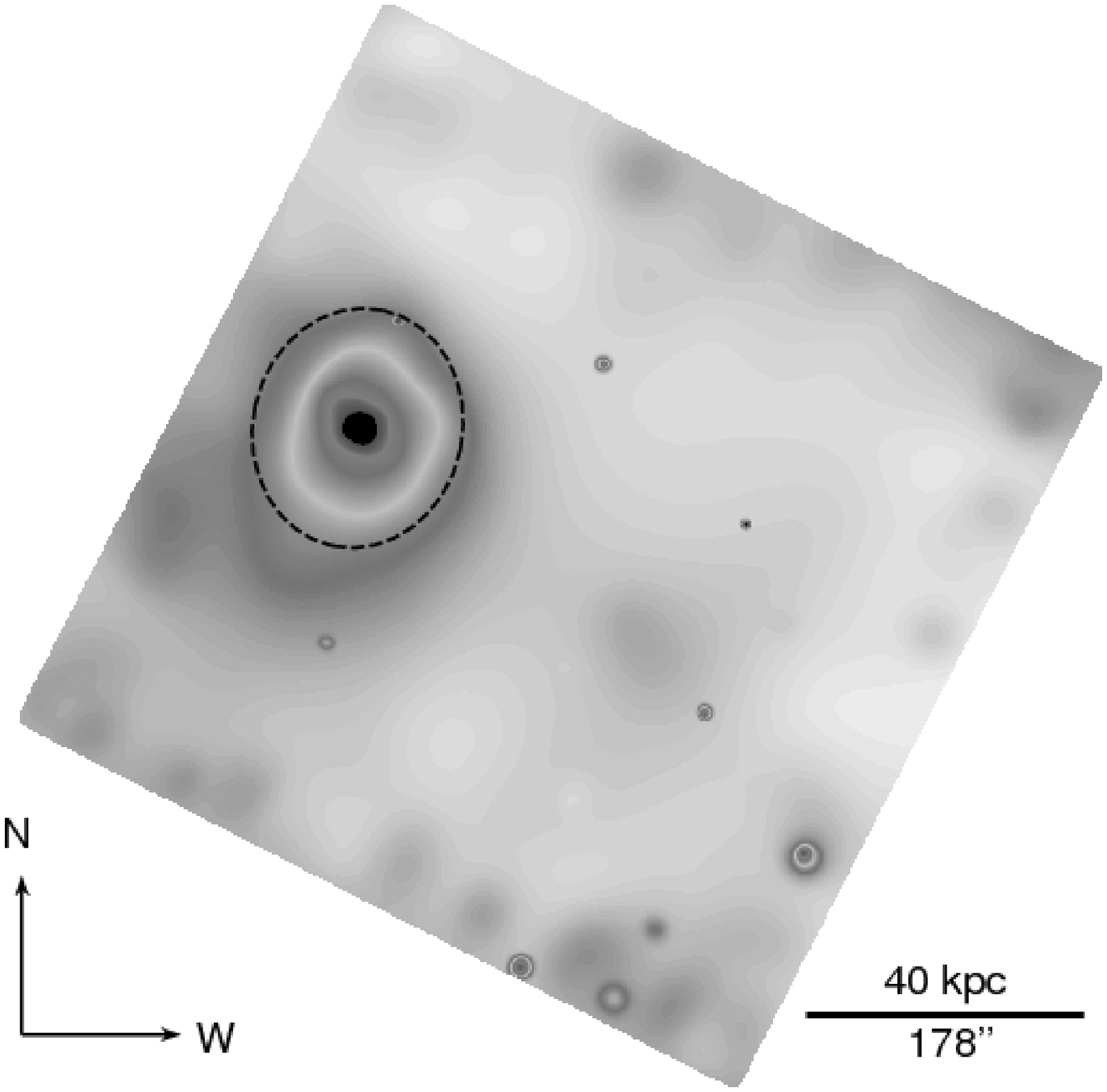,width=6cm}}
}
\caption{\label{fig:smooth}Adaptively smoothed mosaiced 0.3-2.0~keV \xmms\
  EPIC MOS+PN images of NGC~57 (\textit{left}) and IC~1531
  (\textit{centre}), and \chandra\ ACIS-S3 image of NGC~7796
  (\textit{right}). NVSS radio contours marking 3, 9, 27, 81 and 243
  $\sigma$ significance are overlaid on the IC~1531 image.. Dashed black
  lines mark the \Dtf\ ellipse for each galaxy.}
\end{figure*}

Point sources were identified using the \textsc{ciao} task
\texttt{wavdetect} with a detection threshold chosen to ensure $\leq$1
false source in the field. Source ellipses with axes 8 times the standard
deviation of the source distribution were used to exclude the point
sources, with the exception of a source coincident with the optical centre
of IC~1531. Images and spectra were extracted from the cleaned data, and
exposure maps and weighted responses extracted as described in the
\textsc{ciao} threads\footnote{http://asc.harvard.edu/ciao/threads/index.html}.

\textsc{ciao} v3.4 and \textsc{caldb} v3.3 were released during our
analysis of the \chandra\ data. To test the effect of the improved
calibration and handling of serial charge transfer inefficiency (CTI) on
our results, we reprocessed the NGC~7796 observations using the more recent
software and extracted spectra. We found that the correction for serial CTI
increased the number of counts in the energy range 0.3-7.0 keV in the S3
chip by $\sim$1\%. Spectral fitting produced almost identical results, with
gas temperatures changing by $\sim$0.01~keV. We therefore concluded that
our original analysis was satisfactory and unlikely to be significantly
improved by using the improved software.

%\begin{table}
%\begin{center}
%\begin{tabular}{l|l|ccc}
%Galaxy & Dataset & MOS 1 & MOS 2 & PN \\
%\hline
%NGC~57 & Blank field & 0.0376 & 0.0343 & 0.207 \\
% & Telescope closed & 0.174 & 0.220 & 0.825 \\
%\hline
%IC~1531 & Blank field & 0.0366 & 0.0318 & 0.157 \\
% & Telescope closed & 0.162 & 0.197 & 0.771 \\
%\end{tabular}
%\end{center}
%\caption{\label{tab:bgscal} Background scaling factors}
%\end{table}

\section{Results}
Figure~\ref{fig:smooth} shows exposure and vignetting corrected adaptively
smoothed images of the three galaxies. \xmm\ data are used for NGC~57 and
IC~1531, combining data from the MOS and PN cameras. Smoothing was carried
out with the \textsc{sas} \texttt{asmooth} task, with smoothing scales
chosen to achieve a signal-to-noise ratio of 10. NVSS radio contours are
overlaid on the image of IC~1531. The combined ACIS-S3 data are used for
NGC~7796, and smoothing is performed with the \textsc{ciao}
\texttt{csmooth} task with signal-to-noise of 3-5.  All three galaxies have
relatively compact X-ray emission, extending to only a little further than
the \Dtf\ ellipse (which approximates the 25 mag/arcsecond optical
isophote). The surrounding fields show a number of point sources, and there
is also an apparently diffuse feature to the southwest of NGC~57 (marked
with a white cross), which will be referred to hereafter as \xmmu.

\subsection{Surface brightness models}
\begin{figure*}
\centerline{
\epsfig{file=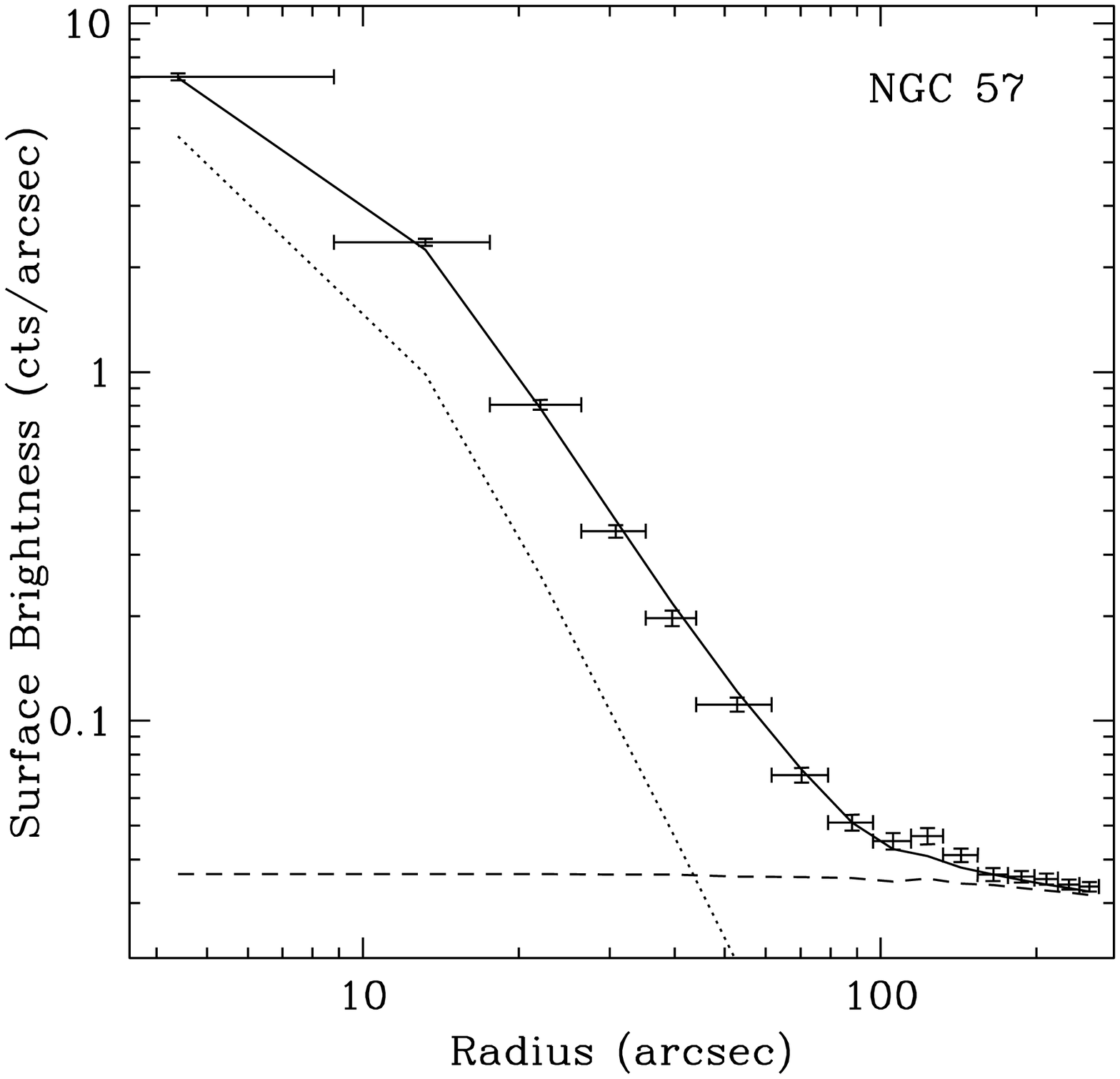,width=6cm,bbllx=20,bblly=210,bburx=570,bbury=750,clip=}
\epsfig{file=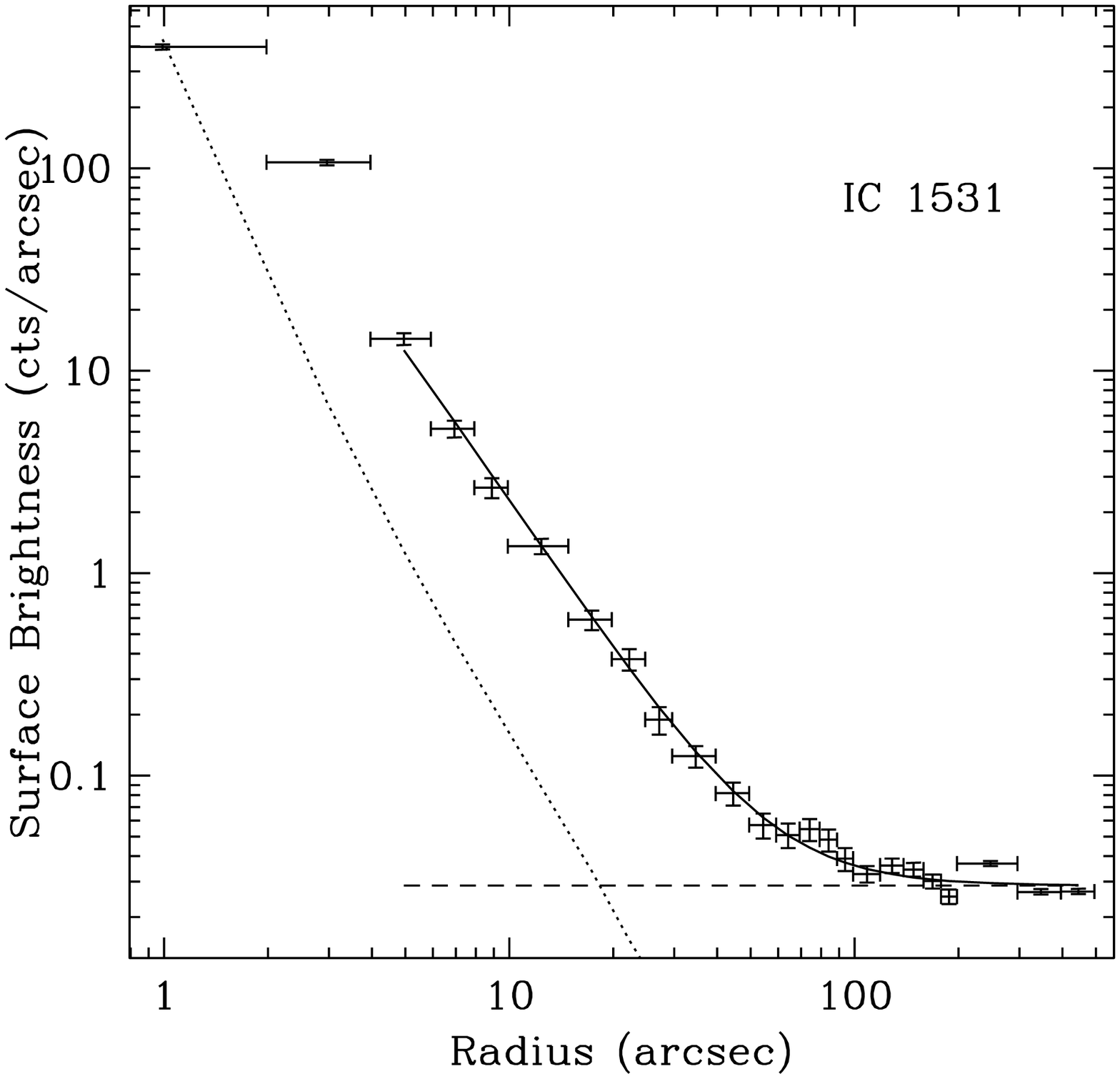,width=6cm,bbllx=20,bblly=210,bburx=570,bbury=750,clip=}
\epsfig{file=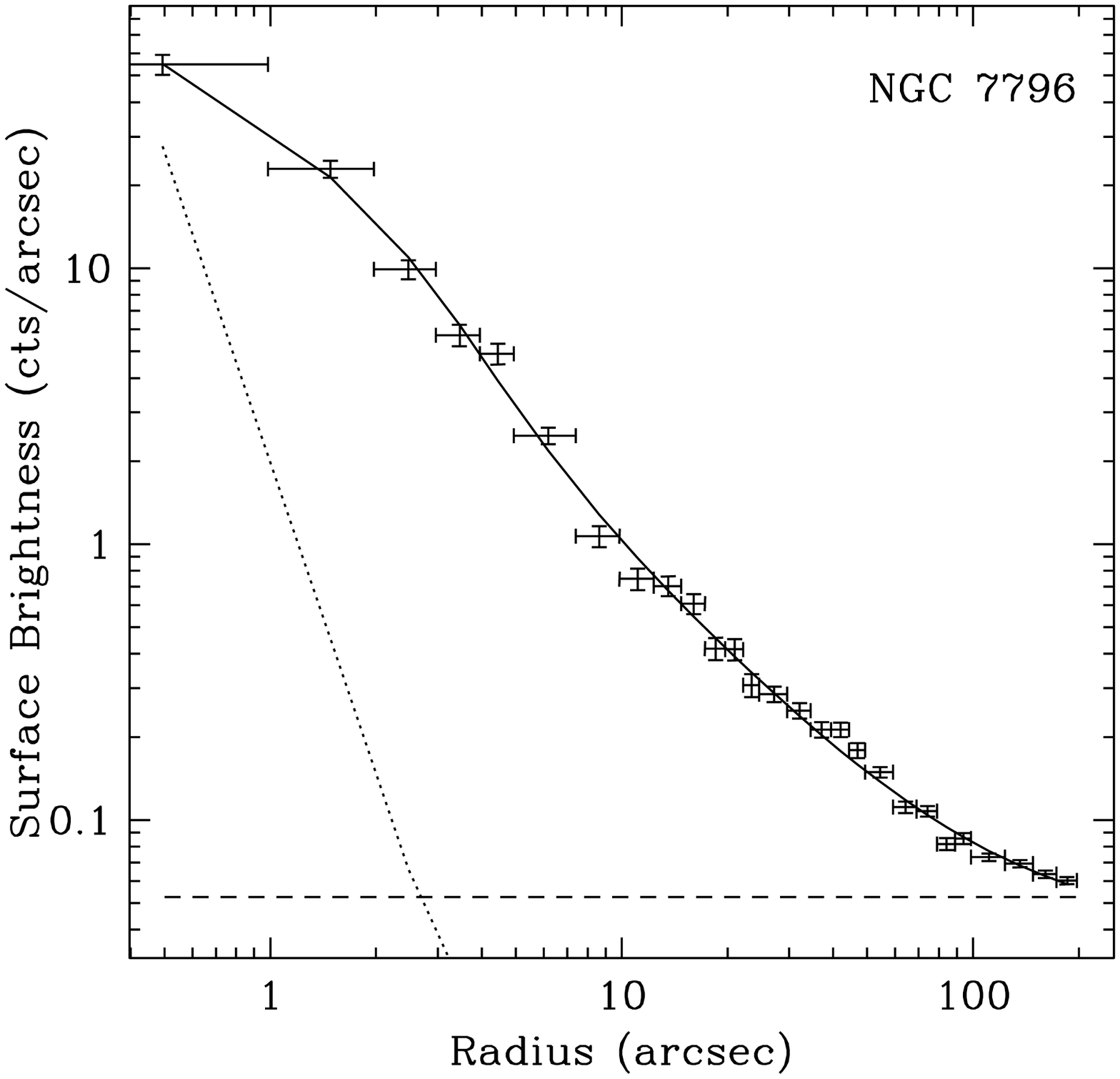,width=6cm,bbllx=20,bblly=210,bburx=570,bbury=750,clip=}
}
\caption{\label{fig:radial}\xmms\ radial surface brightness profile of
  NGC~57 (\textit{left}), and \chandra\ profiles of IC~1531
  (\textit{centre}) and NGC~7796 (\textit{right}), showing the data (error
  bars) best fitting model (solid line) and background (dashed line). The
  \xmms\ and \chandra\ PSFs are marked by dotted lines, but only in the
  case of NGC~57 was a point source component included in the fit. For
  IC~1531, the central two bins were excluded from fitting due to
  contamination by the central point source and jet.}
\end{figure*}

We prepared images of all three galaxies in the 0.3-2.0 keV band, with
point sources removed. The energy band was chosen to maximize the
signal-to-noise ratio for gaseous emission in the sources.  Appropriate
exposure maps were generated for each image. In the case of the \xmms\
data, PSF images were also created and as the number of detected counts
from each camera was relatively low, we summed the \xmms\ images and
associated exposure maps and PSFs from all three cameras to produce a
single set of products for each galaxy. The images were binned to the
EPIC-PN physical pixel size of 4.4\arcs.  For \chandra, the 0.49\arcs\
physical pixel size was used, and we ignored the PSF as it is comparable to
the pixel scale.  Fitting was performed in the \textsc{ciao sherpa}
package. As accurate \xmms\ fitting requires PSF convolution and the
\textsc{sherpa} one-dimensional PSF convolution is known to be inaccurate,
all \xmms\ fits were performed comparing models directly to the images. For
fits to \chandra\ data alone, 1-D fits to the exposure-corrected data are
generally acceptable. As the galaxies occupy relatively small areas near
the focus of each camera, we assumed the background to be flat (after
exposure correction) and included an appropriate model component to account
for its contribution. For 2-D fits, as the source images have many pixels
with few counts (or none), we used the Cash statistic \citep{Cash79} when
fitting. This statistic only provides a relative measure of the goodness of
fit, so that while it allows us to improve fits and find the best solution
for a given model, it gives no information on the absolute goodness of the
final fit. We therefore use azimuthally averaged radial profiles and
residual images to help determine whether particular fits are acceptable.
Figure~\ref{fig:radial} shows radial profiles of our best fits to each
galaxy, and table~\ref{tab:radial} shows the best fitting parameters for
the models used.

\begin{table}
\begin{center}
\begin{tabular}{lccc}
Parameter & NGC~57 & IC~1531 & NGC~7796\\
\hline\\[-2.5mm]
r$_{core}$ (\arcs)& 5.33$^{+0.36}_{-5.33}$ & 1.03$^{+0.33}_{-0.53}$ & 0.35$^{+0.28}_{-0.21}$\\[+1mm]
\Bfit\ & 0.594$^{+0.047}_{-0.109}$ & 0.582$^{+0.009}_{-0.11}$ & 0.390$^{+0.008}_{-0.011}$\\[+.5 mm]
r$_{core,2}$ (\arcs) & - & - & 3.10$^{+26.3}_{-3.1}$\\
\Bfittwo\ & - & - & 0.963$^{+1.046}_{-0.278}$\\
%amplitude & 283.7 & \\
%pnt. src. ampl. & 2693.3 &  
\end{tabular}
\end{center}
\caption{\label{tab:radial} Best fitting beta model parameters and
  1~$\sigma$ errors from the surface brightness fits to the three
  galaxies. The model fit to NGC~57 also included a point source component.}
\end{table}

We found that for IC~1531 the source was only marginally extended in the
\xmms\ data, and therefore difficult to fit accurately. We therefore fit
the \chandra\ data only. IC~1531 hosts a bright radio source \citep[0.48~Jy
at 2.7~GHz,][]{Ekersetal89} which extends $\sim$200~kpc south-east (see
Figure~\ref{fig:smooth}). The \chandra\ data reveals a central X-ray source
with a jet-like extension of a few pixels with a similar orientation to the
radio extension axis. Fitting with the central source included, we found
that the radial profile was best fit by a beta model with no point source
component, as the contribution from the jet mimics an extended central
region. The beta model provided a rather poor fit, with clear deviations
from the data at both large and small radii. Excluding the inner two bins
of the profile, we found that the beta model fit was greatly improved, and
we have taken this as our best fit model for this galaxy. Adding a point
source with normalization fixed to match the central bin produced only very
minor changes in the shape of the beta model (core radius and beta changed
by factors similar to their uncertainties), and only significantly affected
the normalisation of the model.  For NGC~57, we found it necessary to add a
point source model in order to achieve a reliable fit. NGC~7796 is
moderately well fit by a single beta model, but the fit is improved if a
second beta component is added. Comparing the two fits with the F-test, we
find a 0.6 per cent probability that the difference in fit statistic is not
significant, so we consider the two-beta model to be a significant
improvement. We also tried fitting using a beta model with a point source,
but the excess emission above the beta model is clearly more extended than
would be expected from a point source. We therefore accept the two beta
model as the best fit for this galaxy.

\subsection{Spectral models}

We initially fitted the integrated spectra of each galaxy, using circular
extraction regions of radius 150\arcs\ ($\sim$53 kpc) for NGC~57, 30\arcs\
($\sim$16 kpc) for IC~1531 and 100\arcs\ ($\sim$24~kpc) for NGC~7796. The
radii were selected based on the surface brightness fits to contain all the
detected diffuse emission associated with each galaxy.  Nearby point
sources were excluded and, in the case of NGC~57, a region to the southwest
corresponding to the diffuse feature marked in Figure~\ref{fig:smooth}. The
region used to exclude this feature was a circle of radius $\sim$47\arcs\
centred at 00$^h$15$^m$16$^s$.2 +00\degree17\arcm20\arcs. Source and
background spectra were extracted, appropriate responses created, and the
source spectra grouped to 20 counts per bin. The spectra were fitted using
\textsc{xspec} v. 11.3.2z, ignoring energies lower than 0.4 keV and higher
than 7.0 keV. Hydrogen column was held fixed at the galactic value in all
cases. For NGC~57 and IC~1531, we simultaneously fit spectra from all three
\xmms\ EPIC cameras. The \chandra\ spectrum for IC~1531 was fitted
independently as a check on the \xmms\ fits. For NGC~7796 we simultaneously
fitted spectra drawn from the two observations. Experimentation showed that
independent fits to each spectra produced consistent results, though with
large errors in the case of the shorter exposure.

For NGC~57 we found that the spectrum could be fit using a single
temperature APEC plasma model \citep{Smithetal01}, with a powerlaw
component presumably describing emission from unresolved point sources and
the AGN. However, residuals above and below the Fe peak at $\sim$1 keV
indicated that the fit might be improved by the addition of a second plasma
component \citep{Buotefabian98,Buote00b}. This two-temperature model did
produce a somewhat better fit, though with less well constrained errors on
some of the parameters. The results of both fits are shown in
Table~\ref{tab:spectral}. We also tried fits in which individual elemental
abundances were allowed to vary freely, but found no significant
improvement. Similar fits to the \xmms\ data for IC~1531 showed that this
object has a much higher contribution (79 per cent) to its total emission
from the powerlaw component. From the surface brightness models, we can
estimate the fraction of powerlaw emission expected from a point source
with normalisation chosen to match the central surface brightness bin to be
$\sim$85 per cent. Given the uncertainties involved, this seems a
reasonable agreement. Considering its strong radio emission, we believe
that the central AGN is the dominant source of X-ray emission in the
IC~1531. A single temperature fit was found to be acceptable for this
galaxy, and is shown in Table~\ref{tab:spectral}. For NGC~7796, a single
temperature plasma model with a powerlaw component was also found to be
acceptable, and addition of a second plasma component produced no
significant improvement in the fit. However, in this case the plasma
component produces the majority of the flux.
%Examples of the integrated spectra and
%best fit models for both galaxies are shown in Figure~\ref{fig:spectra}. 

\begin{table*}
\begin{center}
\begin{tabular}{lccccccccc}
Galaxy & Model & kT & kT2 & Z & red. $\chi^2$ & d.o.f. & Flux & Fraction$_{PL}$ & $L_{X,gas}$ \\
 & & (keV) & (keV) & (\Zsol) & & & (\ergpspcmsq) & & \\
\hline\\[-2.5mm]
NGC~57 & 1T+PL & 0.93$\pm0.02$ & - & 0.97$^{+0.86}_{-0.44}$ & 1.111 & 346 &
4.710$\times$10$^{-13}$ & 0.60/-/0.40 & 1.78$\times$10$^{41}$ \\
NGC~57 & 2T+PL & 0.83$\pm$0.02 & 1.79$^{+0.54}_{-0.32}$ &
1.31$^{+0.40}_{-0.43}$ & 0.994 & 344 & 4.640$\times$10$^{-13}$ &
0.51/0.26/0.23 & 2.24$\times$10$^{41}$ \\
IC~1531 & 1T+PL & 0.55$\pm$0.03 & - & 0.61$^{+0.11}_{-0.08}$ & 0.895 & 299
& 4.645$\times$10$^{-13}$ & 0.21/-/0.79 & 1.41$\times$10$^{41}$ \\
NGC~7796 & 1T+PL & 0.53$\pm$0.03 & - & 0.53$^{+4.33}_{-0.13}$ & 1.036 & 194
& 3.205$\times$10$^{-13}$  & 0.76/-/0.24 & 7.24$\times$10$^{40}$
\end{tabular}
\end{center}
\caption{\label{tab:spectral} Best fit models to integrated \xmms\ spectra. Spectral models consist of powerlaw (PL) and either single
  temperature (1T) or two-temperature (2T) APEC components. 90\% errors
  are given and fluxes are calculated for the 0.4-7.0 keV band. The
  fraction$_{PL}$ entry shows the fraction of flux produced by each model
  component, in (APEC/APEC/powerlaw) order. All fits were performed with
  \NH\ fixed at the galactic value. }
\end{table*}

As the emission from IC~1531 is largely confined within 30\arcs\ of the
galaxy centre and is dominated by emission from the AGN, we do not attempt
to split the \xmms\ data for this galaxy into radial bins.  With the higher
spatial resolution \chandra\ data we were able to extract spectra from two
regions, a central circular region within 7.5\arcs\ and an outer annulus
with radius 7.5-30\arcs. The central region completely encloses the central
point source and jet structure and we expect it to be dominated by the AGN.
Fitting with an APEC+Powerlaw model, we found a similar temperature for the
plasma component to that measured from the \xmms\ data
(0.55$^{+0.06}_{-0.07}$~keV). Abundance was essentially unconstrained,
probably due to trade-offs against the normalizations of the powerlaw and
APEC models. The powerlaw dominated the emission, producing $\sim$90\% of
the 0.4-7.0~keV flux, for a total luminosity of
2.90$^{+0.15}_{-0.29}\times10^{41}$\ergps, which can be considered as an
upper limit on the X-ray emission of the AGN.  For the outer region, the
gas temperature was marginally lower (0.52$^{+0.05}_{-0.07}$~keV) and
abundance only poorly constrained (0.44$^{+0.51}_{-0.28}$~\Zsol). There is
still a strong powerlaw component in this region, contributing about one
third of the total flux. This probably consists of emission from the X-ray
binary population of the galaxy.

\begin{figure}
\centerline{\epsfig{file=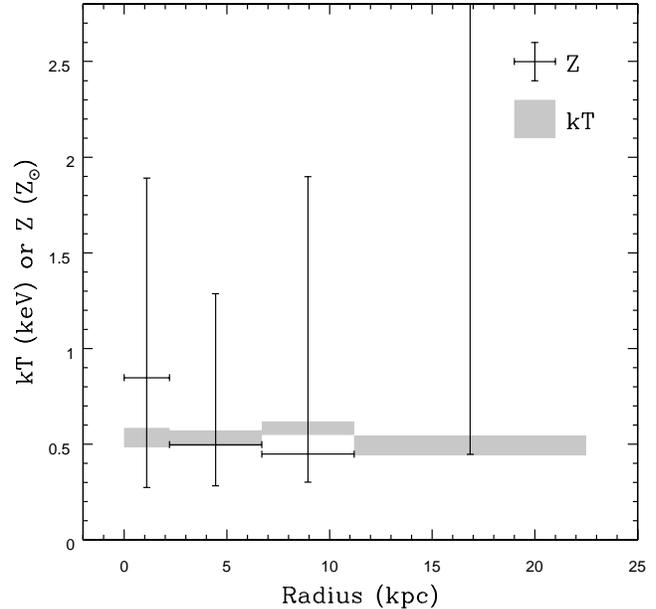,width=8.5cm,bbllx=20,bblly=200,bburx=565,bbury=750,clip=}}
\caption{\label{fig:N7796_kT} Temperature and abundance profiles for
  NGC~7796. 90\% error regions are indicated by the symbols. The abundance
  of the outermost bin is essentially unconstrained.}
\end{figure}

NGC~7796 is more extended than IC~1531 and we were able to split the
emission into four radial bins with outer radii 10\arcs, 30\arcs, 50\arcs
and 100\arcs\ ($\sim$2.5, 7, 12 and 24~kpc). We fitted APEC+powerlaw
models, but found that the powerlaw component was only required in the
inner two bins. Abundance proved to be poorly constrained in all bins,
owing to the trade-offs against the powerlaw component and poor statistics.
However temperature was well constrained and was not affected by leaving
abundance free or fixing it at the value derived from the integrated
spectrum. Figure~\ref{fig:N7796_kT} shows the temperature profile of NGC~7796.
All four bins have best-fitting temperatures of between 0.48 and 0.58 keV,
with errors of 0.05 keV or less. No trend in temperature with radius was
found, and the radial temperature profile agreed closely with the
temperature measured from the integrated spectrum.

\begin{figure}
\centerline{\epsfig{file=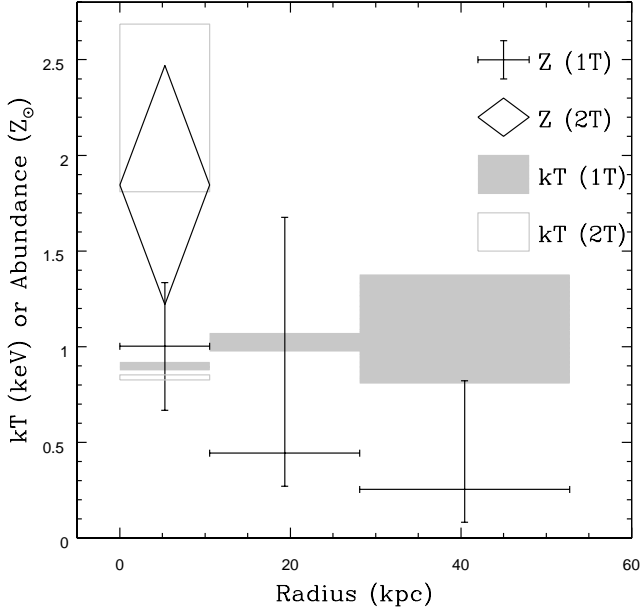,width=8.5cm,bbllx=20,bblly=200,bburx=565,bbury=750,clip=}}
\caption{\label{fig:kTZ} Temperature and Abundance profiles for NGC~57.
  Symbols indicate 90\% error regions on the fitted parameters.}
\end{figure}

NGC~57 has sufficiently extended emission to allow us to extract spectra in
three concentric circular annuli with outer radii 30\arcs, 80\arcs, and
150\arcs\ ($\sim$10, 28 and 53 kpc). We again fitted APEC+powerlaw models,
and results from these fits are shown in Figure~\ref{fig:kTZ}. Fit quality
is similar in all three bins, with reduced $\chi^2$ 0.95-1.15. In the
central bin, this can be improved by the addition of a second APEC model,
which gives reduced $\chi^2$=1.02, The high value of the abundance in the
two-temperature model and its large errors may be an indication that the
data are too poor to support such a complex model rather than an indication
of a real 2-2.5\Zsol\ abundance in the galaxy core. There are indications
of an abundance gradient in the galaxy, but the errors are large enough
that we cannot confirm this. The galaxy halo appears to be approximately
isothermal, with a possible slight fall in temperature in the core. This
could be an indication of radiative cooling in action in the densest parts
of the galaxy halo, but the difference is small, so the assumption of
isothermality is reasonable.

\subsection{\xmmu}
As mentioned previously, there is a small area of apparently diffuse
emission $\sim$2.5\arcm\ to the southwest of NGC~57. The NASA/IPAC
extragalactic Database (NED) contains no object at this position, and the
nearest known extragalactic object is in fact NGC~57. Examination of the
Digitized Sky Survey (DSS) images for this position shows only very faint
sources with too little detail to allow identification.  Comparison with
the PSF of the three \xmms\ telescopes at this off axis angle shows that
the source is definitely extended. In total, \xmmu\ contains $\sim$530
counts, enough for a crude spectral fit. However, the two MOS cameras
contain only $\sim$120 counts each and therefore provide very poor spectra.
We therefore fit models simultaneously to all three cameras, and to the PN
alone. The results of these fits can be seen in Table~\ref{tab:blob}.

\begin{table}
\begin{center}
\begin{tabular}{lcccc}
Parameter & \multicolumn{2}{c}{3 Cameras} & \multicolumn{2}{c}{PN only} \\
 & APEC & Powerlaw & APEC & Powerlaw \\
\hline\\[-2.5mm]
kT (keV) & 1.38$^{+0.61}_{-0.41}$ & - & 0.91$^{+0.58}_{-0.27}$ & - \\[+.6mm]
Z (\Zsol) & 0.01$^{+0.09}_{-0.01}$ & - & 0.01$^{+0.05}_{-0.01}$ & - \\
$\Gamma$ & - & 2.27$^{+0.28}_{-0.26}$ & - & 2.30$^{+0.35}_{-0.31}$ \\
red. $\chi^2$ & 1.19 & 1.16 & 0.873 & 1.107 \\
d.o.f. & 19 & 20 & 10 & 17 \\
Flux ($\times$10$^{-14}$& 2.07 & 2.49 & 2.05 & 2.47 \\
\ergpspcmsq) & & & & 
\end{tabular}
\end{center}
\caption{\label{tab:blob} Spectral fits to source \xmmu.}
\end{table}

These fits are clearly poorly constrained by the data, but it appears that
a powerlaw is a significantly worse description of the PN data. The
abundance of the plasma models is $<$0.06 \Zsol\ at the 90\% level, but it
should be noted that for low temperature sources, fitting a single
temperature model to a source consisting of a multi-temperature plasma is
likely to produce systematically low abundances
\citep{Buotefabian98,Buote00b}.  We therefore consider it fair to
characterise the source as an extended object with a temperature of
$\sim$0.9-1.3~keV, flux of 2-2.5$\times$10$^{-14}$ \ergpspcmsq, and
possibly a low abundance.

The most likely explanations of this source are that it is associated with
NGC~57, or that is a background object which happens to lie close to our
target galaxy. While the temperature of \xmmu\ is similar to that of the
plasma component of NGC~57, the measured abundance is quite different, even
considering the large errors associated with both abundance measurements.
Given the isolation of NGC~57, which suggests that it is unlikely to have
undergone any significant interaction in its recent history, it is
difficult to explain the presence of a large cloud of X-ray temperature gas
separated from the main galaxy halo, with no associated stellar component.
If \xmmu\ were at the same distance as NGC~57, it would be $\sim$20 kpc in
diameter, separated from the galaxy halo by $\sim$30 kpc, and would have a
luminosity of 1.34-1.96$\times$10$^{40}$\ergps. Assuming it to be a sphere
of uniform density (as a first approximation), it would contain
$\sim$3.2-3.5$\times$10$^8$ \Msol.  This is comparable to the hot gas
content of a galaxy.

If \xmmu\ is a background object, we can place constraints on its distance
from the spectral fits. Given the temperature of the system, it would most
likely be a galaxy group or poor cluster. Allowing redshift to vary as part
of the PN spectral fit gives a 90\% upper limit on redshift of $\sim$0.32.
This is equivalent to an angular size distance D$_A$=902.5 Mpc and a
luminosity distance D$_L$=1572.5 Mpc, suggesting that the source is
$\sim$125 kpc in radius and has a luminosity of
6.3-9.2$\times$10$^{42}$\ergps. These values seem quite reasonable for a
large galaxy group.  Alternatively, we can assume an upper limit on X-ray
luminosity of 10$^{43}$\ergps, from which we estimate D$_L < $1980.9 Mpc,
equivalent to a redshift of $\sim$0.39. At this redshift, the system would
have a radius of 140 kpc, again quite compatible with a poor cluster. We
can estimate the apparent magnitude of the dominant galaxy of such a
system, under the assumption that it has a luminosity of
3-10$\times$10$^{10}$ \Lsol, typical of a giant elliptical or cD galaxy.
Using a redshift of 0.32, this is equivalent to an apparent magnitude of
$\sim$19.0-20.2. Although this approaches the limiting magnitude of the DSS
plates, it is perhaps a little surprising that we do not see any more
obvious galaxy candidates.

\subsection{Mass, entropy, and cooling time}
Given the density and temperature distribution of a gaseous halo, it is
possible to estimate a number of three dimensional properties such as mass,
entropy and cooling time. For both NGC~57 and NGC~7796 we can approximate
the temperature structure as isothermal, fitting a constant temperature
model to our radial temperature profiles, and calculate the density profile
based on our surface brightness models. Unfortunately, for IC~1531 we believe
the data are too poor to support such an analysis. The presence of a
powerful radio/X-ray jet source suggests the possibility of interactions
between jet and gas which could mean that the gas is out of hydrostatic
equilibrium. The compactness of the region from which we detect gas
emission ($\sim$16~kpc radius) is not what we would expect for a system
with a relaxed hydrostatic halo. We therefore exclude IC~1531 the following
analysis.

We can use the well known equation for hydrostatic equilibrium,

\begin{equation}
M_{tot}(<r) = -\frac{kTr}{\mu m_pG}\left(\frac{d{\rm ln}\rho_{gas}}{d{\rm
      ln}r}+\frac{d{\rm ln}T}{d{\rm ln}r}\right),
\end{equation}

\noindent to calculate the total mass within a given radius. From density,
temperature and total mass, we can calculate parameters such as gas
fraction, cooling time and entropy where entropy is defined to be

\begin{equation}
S = \frac{T}{n_e^{\frac{2}{3}}}.
\end{equation}

We estimate the errors on the derived values using a monte-carlo technique.
The known errors on the temperature and surface brightness models, and on
other factors such as the total luminosity, are used to randomly vary the
input parameters. We then generate 10000 realisations of the derived
parameters profiles, and use these to calculate the 1$\sigma$ error on each
parameter at any given radius.

\begin{figure*}
\centerline{\epsfig{file=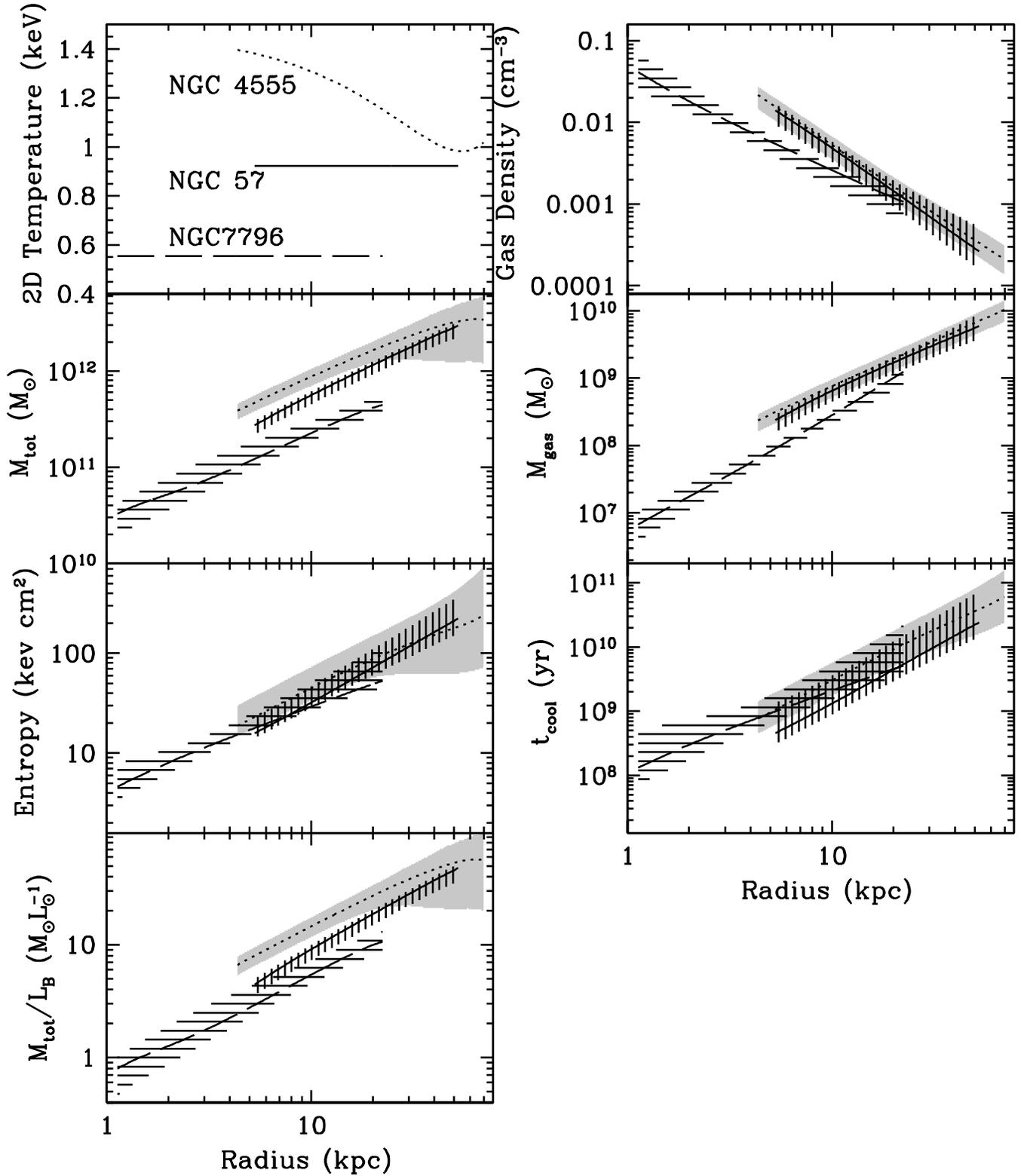,width=18cm,bbllx=22,bblly=110,bburx=554,bbury=730,clip=}}
\caption{\label{fig:mass} Profiles of mass, gas mass, entropy, cooling
  time, etc., for NGC~57 and NGC~7796, with similar profiles for NGC~4555
  plotted for comparison. The dotted line and grey shading show the best
  fit and 1$\sigma$ error regions for NGC~4555, the solid line and
  vertically hatched area NGC~57 and the dashed line and horizontal
  hatching NGC~7796.}
\end{figure*}

We find close agreement between our fits to the radial temperature
profiles and the single temperature fits to the integrated spectra.
Figure~\ref{fig:mass} shows the resulting profiles, overplotted for
comparison on similar profiles for the isolated elliptical NGC~4555
(from OP04). The outer limit of the plots is determined by
the radius to which we can measure the temperature, and the inner limit is
set to be half the width of the central temperature bin. The temperature
profile of NGC~4555 extends further than that of NGC~57 which is itself
more extended than NGC~7796, although we have finer detail in the core of
the latter galaxy.

We note that because we are using an isothermal approximation of the
temperature, the uncertainty of the temperature model is constant with
radius. In the case of NGC~7796 this agrees relatively well with the
results of the radial temperature profile, but for NGC~57 the approximation
does not reflect the larger uncertainty in the outermost temperature bin.
As the total mass is largely determined by the temperature model, the
resulting uncertainty on the total mass for NGC~57 is also constant with
radius. Our choice of assumptions may underestimate the true uncertainty on
total mass; we therefore repeated the analysis using a simple temperature
gradient rather than an isothermal model in order to test whether this
significantly altered the mass profile. The resulting best fitting total
mass is a factor of 1.1 larger, with uncertainties a factor $\sim$1.5
larger.

To calculate the mass-to-light ratio, we use $B$ band effective radii for
the two galaxies \citep[33.1\arcs\ for NGC~57, 25.2\arcs\ for
NGC~7796;][]{Baconetal85,Redaetal05}, assume that the stellar light profile
obeys the standard $r^{1/4}$ law, and normalise the profile by the total
$B$ band luminosity. The assumption of $r^{1/4}$ profiles is supported by
the fits used to estimate the effective radii.  For NGC~57, the
mass-to-light ratio at the inner limit of our calculated profile falls to
$\sim$4, somewhat lower than the usual assumed stellar mass-to-light ratio
of 5-8. However, the total mass-to-light ratio within 55 kpc is
48.38$^{+4.66}_{-10.30}$\Msol/\Lsol.  The mass-to-light ratio of NGC~57 has
been previously estimated from the stellar velocity dispersion in the core
of the galaxy to be 34.40$\pm$0.33 \citep{Baconetal85}, assuming the
stellar orbits to be isotropic. This is comparable with the values we find
at $\sim$30-40 kpc.

NGC~7796 has a lower temperature than NGC~57 and NGC~4555, and a lower
total mass and gas mass at all radii.  The M/L ratio is lower than that
expected for stars alone within $\sim$10~kpc. At the maximum radius to
which we can trace the galaxy halo, M/L is a factor of $\sim$2-3 lower than
that found for NGC~57 and NGC~4555 at the same radius. A M/L ratio profile
based on stellar velocity dispersion measurements \citep{Bertinetal94} has
a flatter slope than our profile and finds M/L larger than our estimate for
radii between $\sim$1-10~kpc. It therefore seems possible that our X-ray
mass estimates do not reflect the true total mass profile for NGC~7796.

Given the differences in gas mass, total mass and temperature of the three
systems, it is notable that they have remarkably similar entropy and
cooling time profiles. Within the uncertainties, the profiles are
indistinguishable over the majority of the range we cover. Cooling time
falls below 10$^9$ yr within 5-10~kpc, suggesting that some form of heating
is required in all cases to prevent excessive cooling. The entropy profiles
all fall to less than 20 keV cm$^{2}$ in the central regions. In NGC~7796
we are able to trace the entropy profile in to $\sim$1~kpc without finding
evidence of flattening in the profile. Central entropy cores have been
observed in some clusters and it has been suggested that they are evidence
that AGN heating provides feedback to balance cooling
\citep{VoitDonahue05}. The lack of such a core in our galaxies may suggest
that no significant outbursts have happened in the recent past, or that the
heating mechanism does not efficiently deposit energy at small radii.

\section{Environment} 
\label{sec:environs}
NGC~57, NGC~7796 and IC~1531 were chosen as examples of early-type galaxies
in very low density environments. The selection criteria were the same as
those used in previous papers \citep[OP04, ][]{Redaetal04}, which specify
that the candidate galaxy should have no neighbours which are:

\begin{enumerate}
\item within 700\kmps\ in recession velocity;
\item within 0.67 Mpc in the plane of the sky;
\item less than 2 $B$-band magnitudes fainter (B$_T$).
\end{enumerate}

Galaxies which are members of a Lyon Galaxy Group \citep{Garcia93} are also
excluded. These conditions are intended to ensure that the candidate
galaxies do not lie in any group or cluster, and that any neighbouring
galaxies are too small or too distant to have had any significant effect on
their evolution or properties. Potentially, these criteria could result in
the selection of fossil groups, systems in which all major galaxies have
merged to form a single elliptical \citep{Ponmanbertram93,Jonesetal03}.
However, our galaxies are too X-ray faint to fall into this category,
having halo luminosities below the minimum value of \Lx=10$^{42}$\ergps\
(see \textsection~\ref{sec:comparison} for further discussion of this
issue).  Selection was performed using data from the NASA-IPAC
Extragalactic Database (NED) and the Lyon-Meudon Extragalactic Data Archive
\citep[LEDA,][]{Patureletal03}, and the results were checked by examination
of Digitized Sky Survey (DSS) images of the region surrounding the
candidates.

\begin{figure*}
\centerline{
\epsfig{file=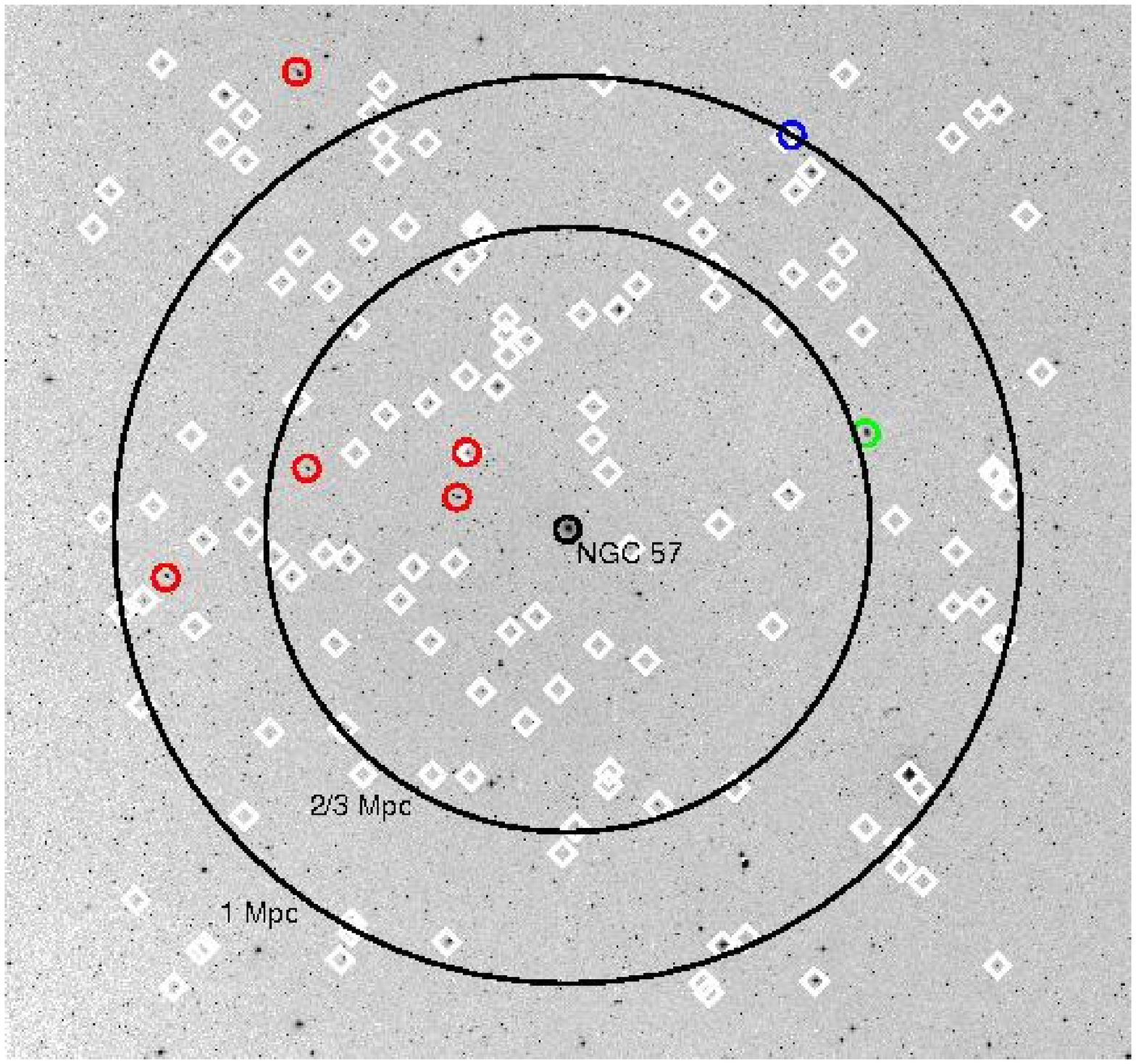,width=6cm}
%\hspace{1cm}
\epsfig{file=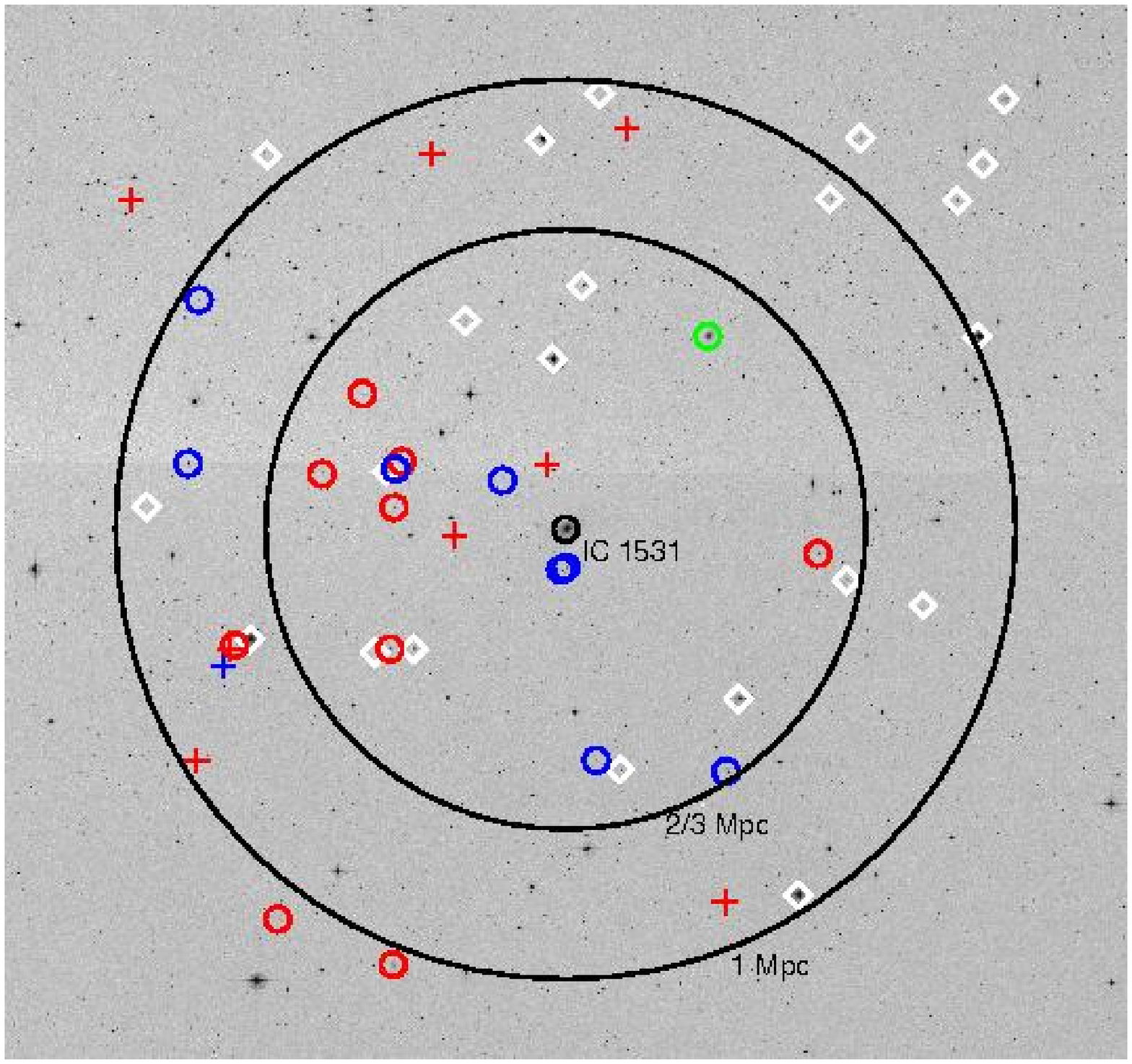,width=6cm}
\epsfig{file=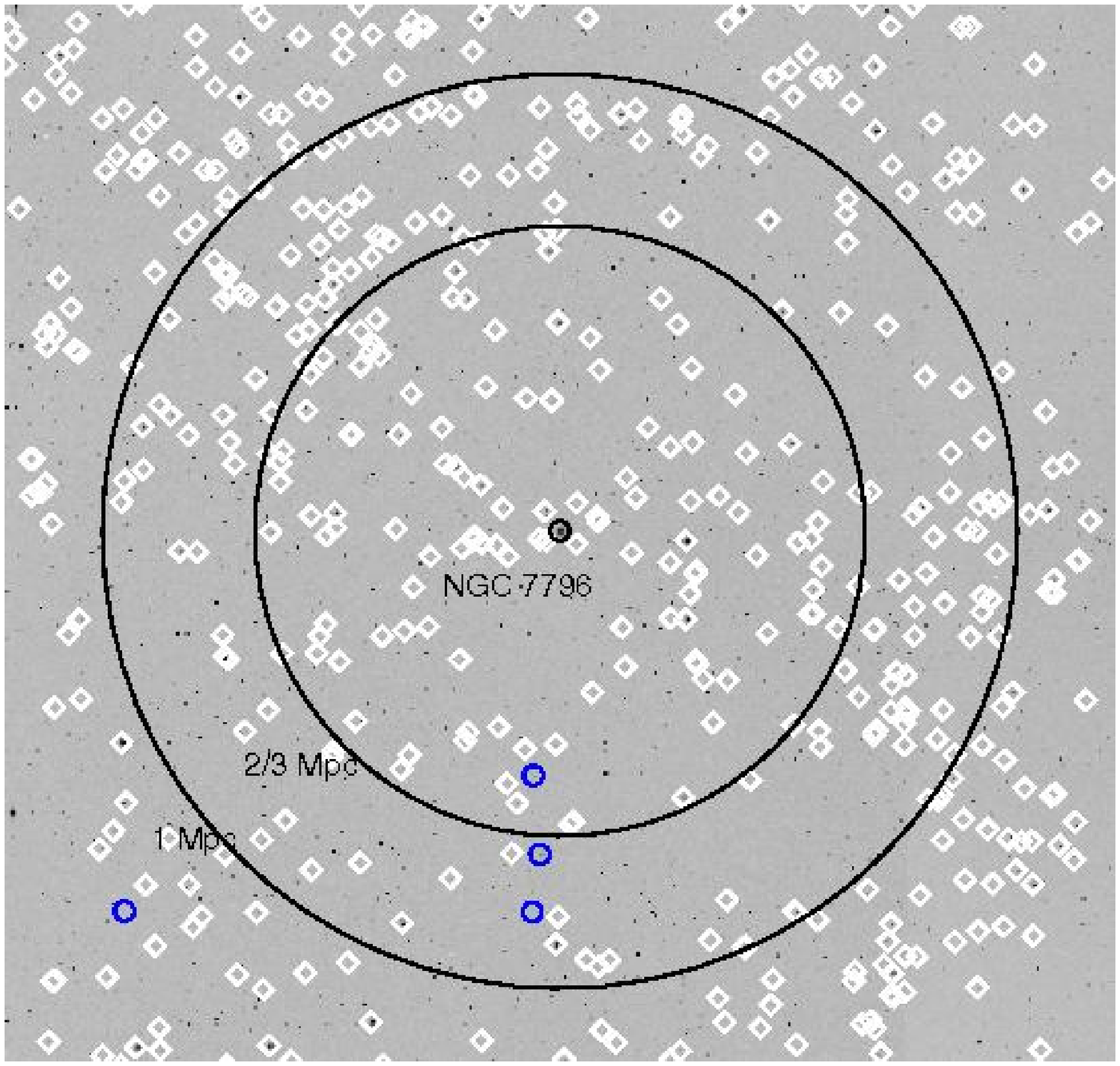,width=6cm}
}
\caption{\label{fig:iso_gals} Digitized Sky Survey images of the regions
  around NGC~57 (\textit{left}), IC~1531 (\textit{centre}) and NGC~7796
  (\textit{right})with the positions of other galaxies marked. Gray
  diamonds mark galaxies with no measured redshift, red points galaxies
  with larger recession velocities and blue points those with lower
  velocities. Green points mark galaxies which are $<$2 magnitudes fainter
  than the target galaxy in B$_T$.  Circles mark galaxies within 700 \kmps\
  of the candidate isolated elliptical and crosses those outside 700
  \kmps\. Circles marking 2/3 and 1 Mpc radius at the assumed distance of
  the candidate are included for scale.}
\end{figure*}

However, since the original selection was performed, both LEDA and NED have
expanded as more galaxies are identified and more information becomes
available. In particular, IC~1531 lies in a 2dF Galaxy Redshift Survey
field \citep[see][for detailed description of the 2dFGRS]{Collessetal01},
which provides us with a much clearer picture of its environment than was
previously possible. NGC~7796 is on the border of another 2dFGRS field.
Figure~\ref{fig:iso_gals} shows DSS images of the three targets, with
neighbouring galaxies marked.

\begin{figure} 
\centerline{\epsfig{file=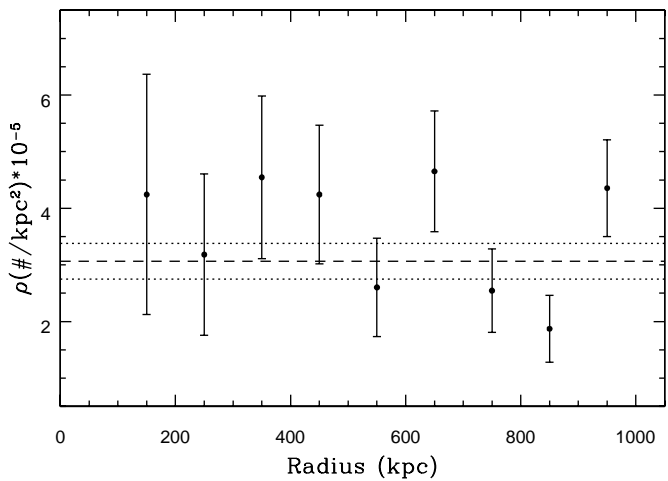,width=9cm,bbllx=50,bblly=370,bburx=550,bbury=740,clip=}}
\caption{\label{fig:iso_histo} Plots of Surface density of galaxies with no
  measured redshift surrounding NGC~57. The points show values in bins 100
  kpc wide (using the assumed distances of NGC~57), with 1$\sigma$ error
  bars. The dashed line shows the best fit constant density model, with
  dotted lines marking the 1$\sigma$ errors. 100 kpc is equivalent to
  4.74\arcm.}
\end{figure}

NGC~57, which does not lie in a redshift survey field, has very few
definite neighbours; only three small galaxies are within 0.67 Mpc and
700\kmps. Between 0.67 and 1 Mpc from NGC~57 are three more galaxies. Two
are small, but the third, IC~4, is a spiral galaxy only $\sim$1.5
magnitudes fainter than NGC~57.  However, the distance between the two is
large enough that NGC~57 still meets the isolation criteria. We have also
tested the possibility that some of the 107 galaxies without redshifts in
the area surrounding NGC~57 might be associated with it, by plotting a
radial surface density profile for these objects, shown in
Figure~\ref{fig:iso_histo}. If there was an unrecognized galaxy group
surrounding NGC~57, we would expect to see a high surface density of
galaxies at small radii, falling to lower values at larger distances. We
find no such trend, which suggests that there is no larger structure around
NGC~57. It should be noted that we have not applied any magnitude cut when
selecting the surrounding galaxies. The lack of such a limit makes a direct
comparison with similar studies of galaxy surface density
\citep{Smithetal04,Redaetal04} impractical.  We cannot consider our sample
complete, and it is possible that inhomogeneities in the coverage of the
field could suppress the signs of structure around NGC~57. However,
examination of the DSS image suggests that we are not missing any bright
galaxies in the field and that any additional neighbours must be relatively
low luminosity galaxies.  We therefore believe this elliptical to be
truly isolated.

The 2dFGRS field in which NGC~7796 falls covers the lower left quadrant of
the area shown in Figure~\ref{fig:iso_gals}. The redshift coverage around
the galaxy is therefore very uneven. However, only three galaxies within
1~Mpc on the sky are found to have redshifts within 2000\kmps, and all
three are more than 2 magnitudes fainter in $B$.  Considering the galaxies
with no redshift measurements, we again plot the radial surface density in
Figure~\ref{fig:iso_histo}. The surface density appears to be approximately
flat out to 1~Mpc, with the exception of the region within 100~kpc of
NGC~7796, where we find 7 galaxies. The overdensity suggests that some of
these may be satellite galaxies associated with NGC~7796. However, they are
very faint; the brightest is 4.2 magnitudes fainter than NGC~7796,
equivalent to \LB=7.7$\times10^8$~\LBsol\ at the distance of NGC~7796. We
therefore conclude that while NGC~7796 may have a small number of companion
dwarf (\LB$<10^9$~\LBsol) galaxies, they are probably too small to have had
any significant impact on the formation history of the galaxy. There is no
evidence of a group or an extended halo of galaxies around NGC~7796.

\begin{figure}
\centerline{\epsfig{file=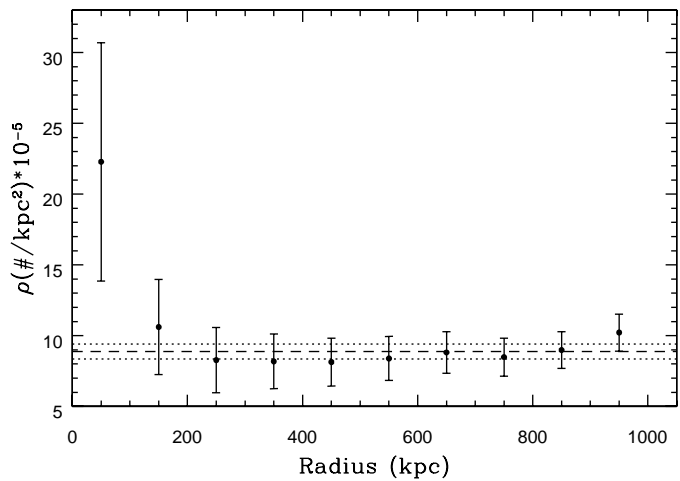,width=9cm,bbllx=50,bblly=370,bburx=550,bbury=740,clip=}}
\caption{\label{fig:iso_histo_N7796} As Figure~\ref{fig:iso_histo}, for the
  galaxies with no measured redshift around NGC~7796. 100~kpc is equivalent
  to 6.89\arcm.}
\end{figure}

IC~1531 is a slightly more difficult case. The 2dFGRS data show there to
be a relatively large number of small galaxies in the surrounding area, 13
within 700\kmps\ and 0.67 Mpc, 24 within 2000\kmps\ and 1 Mpc. One of
these galaxies, PGC132744 (also known as 2MASX J00084885-3203089) is an Sc
galaxy whose B$_T$ magnitude in LEDA is 15.49, only 1.65 magnitudes fainter
than IC~1531. This means that IC~1531 fails the initial criteria for
isolation. However, both galaxies have near infrared magnitudes available,
and in J, H and K$_S$ bands, PGC132744 is 3.9-4.3 magnitudes fainter than
IC~1531. PGC132744 also appears to have a rather clumpy morphology, and its
extremely blue colors (B-K$_S$ = 5.94) suggests that it may be undergoing
a burst of star formation or some other disturbance which has increased its
B-band surface brightness. We therefore conclude that IC~1531 should not be
ruled out as an isolated elliptical on the basis of this object alone. 

\begin{figure}
\centerline{\epsfig{file=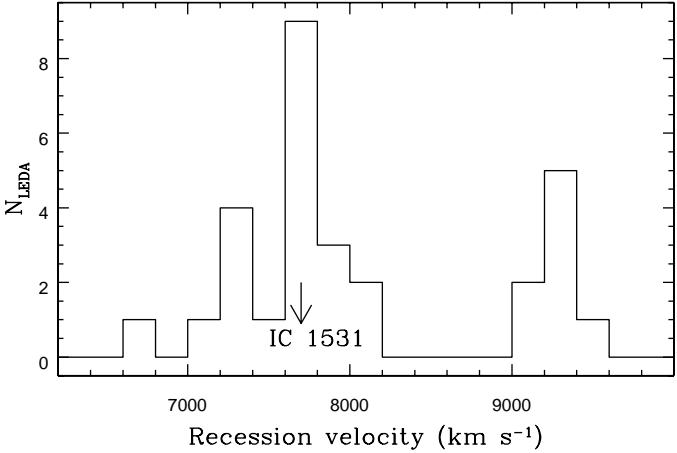,width=9cm,bbllx=40,bblly=370,bburx=550,bbury=730,clip=}}
\caption{\label{fig:vel_histo} Redshift distribution of the galaxies within
  1 Mpc (in the plane of the sky) of IC~1531. The arrow shows the redshift
  of IC~1531.}
\end{figure}

Figure~\ref{fig:vel_histo} shows the distribution of recession velocities
for the galaxies in the neighborhood of IC~1531. The galaxies appear to be
arranged in two velocity regions, either within $\sim$1000\kmps\ of IC~1531
or in a smaller clump at $\sim$9300\kmps. However, examination of
Figure~\ref{fig:iso_gals} shows that the galaxies at larger recession
velocities are very scattered in the plane of the sky, suggesting that some
of them are very distant. Ignoring these objects, we calculate the mean
recession velocity of the remaining galaxies to be $\sim$7600\kmps, only
100 \kmps\ from the recession velocity of IC~1531, with a velocity
dispersion of line-of-sight 338 \kmps. With the exception of PGC132744, the
galaxies in this velocity grouping have luminosities of 8.5$<$log
\LB$<$9.6, so that the brightest of these neighbours is a factor of
$\sim$18 less luminous than IC~1531, and presumably correspondingly less
massive.  If we assume that these galaxies form a bound system, we can also
estimate the total system mass, using their positions and velocities. We
use the Virial theorem and projected mass methods described in
\citet{BahcallTremaine81}, and obtain masses of
$\sim$2-4.5$\times$10$^{15}$\Msol. This is comparable to a massive cluster.
From the Mass-Luminosity and Mass-Temperature relations measured for galaxy
clusters, we would expect a system of this mass to have an X-ray luminosity
of $\sim10^{45}$\ergps\ and a temperature $\sim$10~keV
\citep{ReiprichBohringer02,Sandersonetal03}. Clearly this strongly
indicates that the system is not truly bound. It seems more likely that
IC~1531 lies in a region where a galaxy group will form in future, perhaps
as part of a filament between larger scale galaxy structures.

\section{Discussion} 
\label{sec:discuss}
As discussed in the previous section, all three of the galaxies observed
appear to be in quite sparse environments, with much lower galaxy densities
than those found in the cores of groups and clusters. Our examination of
the environment of NGC~4555 (in OP04) found it to in an environment similar
to that of IC~1531. Both galaxies are isolated by our criteria, and are not
part of any larger virialized system, but have a sufficient number of low
luminosity neighbours that it seems likely that they are in regions where
groups may form in future. NGC~57 and NGC~7796 are even more isolated, with
only a handful of small neighbours. The very low densities of their
environments suggest that they may never become part of a galaxy group. In
all cases it seems very unlikely that the development of our target
galaxies has been affected by interactions with neighbouring objects. We
therefore feel secure in considering them as isolated, individual systems.
Given the similarity of their environments, and therefore the possibility
of similarity in their formation histories, the variation in their
properties is interesting.

\subsection{Galactic winds vs. hydrostatic haloes}
NGC~4555 and NGC~57 both have extended X-ray haloes, detectable to
$\sim$50-60~kpc, containing $\sim$10$^{10}$~\Msol\ of gas within this
radius, and with temperatures of \gtsim1~keV. M/L ratio profiles for these
two galaxies fall to values similar to those expected for stars alone (5-8
\ML) at small radii, but there is no strong indication of deviations from
hydrostatic equilibrium. Given the short cooling times in the cores of the
two galaxies, and the evidence from surface brightness and spectral fitting
of central point sources and powerlaw spectra, there is the possibility
that cooling is being balanced by low-power AGN activity. In this case we
might expect some deviation from hydrostatic equilibrium. However, seems to
be a relatively minor effect, and we consider it unlikely that our
estimated mas profiles for these two systems have been significantly
effected, particularly at large radius.

In contrast, NGC~7796 and IC~1531 both have somewhat cooler ($\sim$0.5~keV)
and in the case of IC~1531 more compact haloes, and the estimated total M/L
ratio profile of NGC~7796 drops below the expected stellar M/L ratio within
$\sim$10~kpc.  Given the disagreement with the Bertin et al. M/L profile
based on stellar velocity dispersion measurements, this suggests that the
halo is not in hydrostatic equilibrium, at least in the central part of the
galaxy. One possibility is that the gas in NGC~7796 is in a wind state,
with supernovae heating the gas and producing significant motions
\citep[e.g.][]{Pellegriniciotti98,Ciottietal91}. Galaxy winds can have a
range of structures depending on parameters such as SN heating rate, galaxy
mass and density profile. In the case of NGC~3379, an elliptical where the
X-ray mass estimate is observed to be low compared to dynamical estimates,
wind models have been applied to show that a galaxy wind can produce a mass
underestimate of a factor $\sim$2 within the effective radius
\citep{PellegriniCiotti06}. An underestimate of a factor of 2-3 is required
to bring our X-ray mass measurements into line with the dynamical estimate.
The 0.5~keV temperature of NGC~7796 is quite typical of that expected from
a wind system. It is also notable that the modelling of NGC~3379 shows that
the mass discrepancy rises to a peak at small radii, perhaps providing an
explanation for the low values of M/L we observe in the inner regions.
Without detailed modelling of NGC~7796 it is difficult to judge the
reliability of the mass profile.

The presence of an AGN and extended radio source in IC~1531 is interesting,
in that it raises the question of interaction between the AGN jet and the
gaseous halo.  We estimate the total radio power of the source to be
L$_{Radio}$=6.6$\times10^{41}$\ergps\ (10~MHz-10~GHz) based on the
measurements of \citet{Ekersetal89}. The X-ray luminosity is similar,
$\sim$2.9$\times10^{41}$\ergps in the 0.4-7.0~keV band. The radio
luminosity is comparable to that of some cluster central galaxies; M87 has
L$_{Radio}$=7.5$\times10^{41}$\ergps. We can roughly estimate the
mechanical power output of the radio jets from the relation of
\citet{Birzanetal04}, and find L$_{mech}$=1.24$\times10^{43}$\ergps.
Comparison with the luminosity of the gaseous halo shows that $\sim$1\% of
this energy would be sufficient to counteract radiative cooling. This
suggest that either IC~1531 is in the process of losing its halo through
AGN heating and evaporation (this should take no more than a few
10$^7$~yr), or that the jets have cleared channels through the halo we
observe and now deposit very little energy in the gas. 

Based on the global spectral fit and surface brightness modelling of IC~1531
we estimate the gas mass within 30\arcs\ of the galaxy core to be
5.96$^{+6.07}_{-2.16}\times10^8$~\Msol\ (1$\sigma$ uncertainties). The
large errors on the best fitting value are driven by the uncertainty in the
relative normalisations of the plasma and powerlaw models. While the upper
end of this range is comparable with NGC~4555 and NGC~57, the best fitting
value is similar to that found in NGC~7196 at the same physical radius.
Given the compactness of the halo, it seems likely that the AGN activity
has expelled some gas from the system already and that the halo is not in
undisturbed hydrostatic equilibrium. It is possible that IC~1531 also hosts
a supernova driven galactic wind. The time taken to accumulate the observed
gas mass can be estimated based on a present day stellar mass loss rate
from asymptotic giant branch stars of
\.{M}$_*=0.078(L_B/10^{10}L_{B\odot})~M_\odot~yr^{-1}$, comparable to the
stellar mass loss rates derived from \textit{Infrared Space Observatory}
observations of nine ellipticals \citep{Atheyetal02}. From this rate, we
estimate that $\sim$10$^9$~yr are required to build up the current gas
content. We estimate the cooling time in the outer part of the halo to be
$\sim$3$\times$10$^9$~yr, suggesting that stellar mass loss is introducing new
gas faster than cooling can remove it.

AGN heating could also be partially or largely responsible for driving such
a wind. As mentioned above, only a few per cent of the likely energy output
of the jet would be sufficient to significantly heat the gas and
potentially produce a global outflow.  Alternatively, the halo might have
been truncated by AGN heating.  It has been suggested that in some systems
energy may be deposited at the point where jets cease to be tightly
collimated, leading to the heating and removal of the outer part of the
halo while the inner region remains unaffected
\nocite{Sunetal05a,Sunetal05b}(Sun et al. 2005a,b). It is possible that
IC~1531 has lost gas in this way, with the jets stripping away the outer
halo but leaving some gas in the inner core. Jet/gas interactions might be
expected to produce structure in the halo via shocks or compression of the
gas. We see no indication of such structures, or of strong asymmetries in
the halo, but we note that outside the central 7.5\arcs\ where the X-ray
jet dominates, only $\sim$300 counts (0.3-1.5~keV) are detected by
\chandra, making assessment of the halo morphology difficult. Higher
resolution radio observations might provide further indications of the
current relationship between the jet and gas, and show over what range the
jets remain collimated.

\subsection{Comparison with mass estimates for other early-type galaxies}
\label{sec:comparison}
As mentioned in \textsection~\ref{sec:intro}, the dark matter content of
elliptical galaxies is currently a subject of debate. A number of
mechanisms have been used to estimate mass profiles in ellipticals, but
each suffers from potential issues which can affect the accuracy of the
result. Gravitational lensing, perhaps the most direct way of measuring
mass, is capable of tracing the mass profile out to $\sim$100$r_e$
\citep[e.g.][]{Gavazzietal07}. However, it relies on stacking multiple
galaxies to produce the required signal-to-noise and so provides
information on the average properties of the population rather than on
individual systems and the differences between them. 

Kinematic modelling using stellar, globular cluster (GC) and planetary
nebula (PN) velocities provides another mechanism for measuring mass
profiles. Mass measurements based on stellar velocity dispersions are
limited to relatively small radii by the decreasing surface brightness of
the stellar population, but the inclusion of PNe or GCs can extend profiles
out to $\sim$5-10~$r_e$. Unfortunately a number of factors can affect the
accuracy of the kinematic models used. For example, models of merging disc
galaxies suggest that the merger process may produce a population of stars
on very radial orbits \citep{Dekeletal05}. In the early-type galaxy which
results from the merger, this population of stars will produce a large
number of the PNe observed at large radii, introducing a bias which must be
accounted for in any mass models. It has been suggested that early results
based on PNe modelling which found low DM content \citep{Romanowskyetal03}
could be affected by this problem, though it may not explain all of the
discrepancy between observations and predictions
\citep{Dekeletal05,MamonLokas05b}. The large numbers of tracers (GCs or
PNe) required to characterise the anisotropy of the velocity distribution
renders this technique difficult for many systems, particularly those with
sparse GC/PN populations.

\citet{Napolitanoetal05b} provide a collection of mass profiles based on
kinematical modelling in 21 nearby ellipticals. They find the typical M/L
ratio at 0.5$r_e$ to be $\sim$7\ML, providing a good measure of the typical
M/L of the stellar population with minimal DM contribution, though
in some systems values as low as $\sim$3 are reported.  The outer radius to
which the M/L ratio can be traced varies with the study, and there is
considerable variation between galaxies, with some systems showing no
evidence of an increase in M/L (e.g. NGC1379, NGC4697). There is some
indication that the most optically luminous galaxies are also the most dark
matter dominated, with some fainter ellipticals showing little evidence for
massive DM haloes. The authors suggest that this may be evidence
for haloes with lower central concentrations than predicted by standard
$\Lambda$CDM models.  For comparison with our isolated ellipticals we
consider the M/L ratio at $\sim$5$r_e$. This is the approximate limit of
our profiles, and should be large enough to ensure that the dark halo is
the dominant mass component, rather than the stellar population. The
Napolitano et al. sample suggests that a M/L ratio of 20-30\ML\ at this
radius is fairly typical of the more luminous galaxies. The Napolitano et
al sample contains only one galaxy of comparable isolation to the three
systems discussed in this paper, NGC~821, which has a M/L ratio of
13.1$\pm$3.9 \ML\ at 4.8$r_e$. Unfortunately it has almost no hot gas
content\citep{Pellegrinietal07}, so a direct X-ray comparison is not possible.

The number of available mass estimates from X-ray data has improved greatly
since we reviewed this area in OP04. The most notable additions are the
studies of \citet{Fukazawaetal06}, who analysed data for 53 elliptical
galaxies drawn from the \chandra\ and \xmms\ archives, and
\citet{Humphreyetal06b} who performed a detailed study of seven ellipticals
chosen to have relaxed haloes. Fukuzawa et al. report values of 5-80\ML\ at
5$r_e$ with variations of at least one order of magnitude between galaxies
at all radii. Their M/L ratio profiles show a correlation with the size and
luminosity of the X-ray halo, with some faint, compact systems showing M/L
ratios as low as 2~\ML. They also find an increased scatter in the profiles
at small radius. Humphrey et al. find M/L ratios of $\sim$7-25\ML\ at
5$r_e$.  All seven systems have M/L profiles which rise with radius,
indicating the presence of a dark halo.

As is to be expected, there are issues which require consideration when
interpreting these results. The association of low X-ray luminosity with
systems of low M/L ratio, found by Fukuzawa et al., could be be explained
by those galaxies hosting galactic winds rather than hydrostatic haloes. If
this is the case, then the low M/L values found for the compact systems
would be incorrect, owing to an underestimation of the total mass.  The
large variation in M/L at small radii may be the product of AGN heating,
which is likely to have the largest effect on the pressure and temperature
of the interstellar medium (ISM) in the galaxy core. Although Humphrey et
al.  selected their galaxies to have relaxed haloes, it is notable that
they find their lowest M/L ratio, highest stellar M/L ratio, highest virial
mass and lowest DM halo concentration for NGC~4261, an FR-I radio galaxy
with radio and X-ray jets which extend to \gtsim3.5$r_e$
\citep{Zezasetal05}.  Fitting the mass profile with a model consisting of
stellar and DM components, Humphrey et al. find that the stars dominate the
mass profile out to $\sim$6$r_e$ ($\sim$20~kpc). While it is possible that
the galaxy does indeed have such a low density of dark matter in its inner
regions, it also seems possible that jet heating and disturbance of the
inner halo could have affected the estimated mass profile. The two samples
also contain galaxies from a range of environments, ranging from field
galaxies to those in nearby clusters and a large number of group dominant
ellipticals. While this may not affect the mass profile within a few $r_e$,
it suggests that caution should be employed when considering results at
large radii, where contributions from the dark matter halo of any
surrounding system need to be considered.

The Humphrey et al. and Fukuzawa et al. samples also contain two fairly
isolated galaxies, NGC~720 and NGC~6482. The latter was originally selected
as a fossil group candidate, and \chandra\ observations show the halo to be
luminous and extended, providing a degree of support for this
classification \citep{Khosroshahietal04}. However, Buote et al. argue that
the total system mass is significantly lower than groups such as NGC~5044
and that NGC~6482 should be considered as an isolated elliptical.
Unfortunately a detailed study of the galaxy population around
NGC~6482, which might resolve the issue, is not yet available, making a
definitive classification difficult. NGC~720 is also a borderline object;
it has been identified as the dominant elliptical of the LGG~38 group
\citep{Garcia93} but the other members of the group are all considerably
fainter. NGC~720 would meet our selection criteria for isolated galaxies
were it not for its prior identification as a group.Buote et al. argue that
the mass of the system is too low to represent a galaxy group, and its
reported X-ray luminosity falls below the minimum required for fossil
groups. However, an examination of the local surface density of galaxies
with no measured redshift (as performed for NGC~7796 and NGC~57 in
\textsection~\ref{sec:environs}) suggests that there may be a significant
excess of galaxies around NGC~720. This could indicate that the galaxy is
in fact part of a fossil group-like system, but more detailed studies are
required for a definitive judgment. For comparison with our galaxies, we
note that both systems may fall in the category of isolated ellipticals,
but that both appear to be in rather denser environments than any of our
targets.

Our mass profile for NGC~57 only extends to 4.75$r_e$, at which point the
M/L ratio is 44.7$^{+4.0}_{-8.5}$~\ML\ (1$\sigma$ uncertainties). From OP04
the value for NGC~4555 at 5$r_e$ is 43.2$^{+15.2}_{-21.6}$~\ML\ and for
NGC~7796 (assuming hydrostatic equilibrium) 10.6$^{+2.5}_{-2.3}$ \ML. It is
clear that at this radius, NGC~57 and NGC~4555 are quite similar, and are
fairly massive systems with considerable DM content. Their M/L ratios are
rather higher than the dynamical estimates given by Napolitano et al. for
systems of similar $B$-band luminosity. However, the uncertainties on our
estimates are large enough that the differences are not strongly
significant, particularly in the case of NGC~4555. In any case the mass
profile of NGC~57 strongly indicates the presence of a fairly massive dark
halo, lending weight to the result of OP04 that individual ellipticals can
possess significant quantities of DM regardless of environment.

Humphrey et al. suggest that NGC~4555 may possess a group-scale dark halo
and have formed through a process similar to that of a fossil group.  The
velocity dispersions of NGC 4555 \citep[350\kmps,][]{Wegneretal03} and
NGC~57 ($\sim$326\kmps, LEDA) are rather higher than those of IC~1531 and
NGC~7796 (226 and 260\kmps\ respectively). Both NGC~57 and NGC~4555 fall at
the upper end of the \sigT\ relation for ellipticals
\citep{Fukazawaetal06,OPC03}, having parameters comparable to many group
dominant ellipticals. This could be an indication of the presence of
group-scale dark matter halos. From a technical viewpoint, both NGC~4555
and NGC~57 fail to meet the criteria for classification as a fossil group.
Although both galaxies meet our isolation criteria, neither is quite as
isolated as is usually required for a fossil group (no neighbours less than
2 magnitudes fainter within 2000\kmps\ and 1~Mpc). However, this does not
rule out the possibility that these could be galaxies in a group-sized DM
halo which have formed by multiple mergers at an early epoch.  Both
galaxies are fainter than the minimum luminosity of 10$^{42}$\ergps\ used
to select fossil groups, and neither X-ray halo is as extended as would be
expected for a group-scale object. It is difficult to see how a group-scale
system which formed rapidly at an early epoch could have such a low X-ray
luminosity. We would expect a halo with a X-ray luminosity and gas content
similar to any normal group; as there is no evidence of excessive star
formation, we would have to assume that the group had somehow lost
$\sim$80-95 per cent of its gas content through evaporation, a very extreme
scenario. This argument does not rule out a fossil group-like formation
mechanism, though it does suggest that the total mass of these galaxies is
likely to be considerably lower than that of such groups.

More detailed mass modelling, which given the distance of these two systems
requires deeper X-ray observations, could provide insight into the
relationship of these galaxies to fossil groups. Similarly, measurement of
the local galaxy luminosity functions could determine whether NGC~4555 and
NGC~57 have the large population of gravitationally bound satellite
galaxies expected in a fossil group. An argument against a fossil
group-type formation history for theses systems is the relatively flat
galaxy surface density profiles shown in Figures~\ref{fig:iso_histo} and
\ref{fig:iso_histo_N7796}, which suggest the presence of only a few small
neighbouring galaxies. However, better optical data is required if this
issue is to be resolved. Perhaps a stronger argument that the formation
mechanism of isolated ellipticals is not the same as that of fossil groups
is that a significant number of systems studied in the optical show signs
of recent mergers or interactions \citep{Redaetal04}.  However, further
investigation of this area is required to determine how isolated
ellipticals form, and whether there is a continuum of properties, with
fossil groups as the more luminous counterparts of the systems considered
here.

The lower M/L found for NGC~7796 suggests that it is considerably less
massive than NGC~57 and NGC~4555, if we assume it to be in hydrostatic
equilibrium. It is comparable with several of the ellipticals with lower
optical luminosities in the Napolitano et al. sample, including NGC~821,
which is similarly isolated. This emphasizes the difference between
NGC~7796 and NGC~57, which has a near-identical optical luminosity. A
number of possibilities could explain the low value found for this system.
If we assume that the mass profile we measure is accurate, at least in the
outer part of the galaxy, then we must assume either that NGC~7796 has only
a minimal dark matter halo, or that its halo has a relatively large core
and that we do not trace the mass profile far enough to see the rise in M/L
ratio. The latter scenario could suggest either a deviation from the
expected NFW DM profile, as is suggested by Napolitano et al. to explain
the lower mass systems in their sample, or possibly a low concentration NFW
halo, as found by Humphrey et al. for NGC~4261. However, the very low M/L
ratio found in the centre of NGC~7796 suggests that our mass profile may
well be unreliable in the galaxy core, and in this circumstance we must
consider the possibility that the halo as a whole is not in hydrostatic
equilibrium, but is in a galactic wind state.

\section{Summary and Conclusions}
\label{sec:conclusions}
We have used \chandra\ and \xmm\ to observe three isolated early-type
galaxies, NGC~57, IC~1531 and NGC~7796. All three appear to be in very low
density environments with few neighbouring galaxies of comparable mass.
None of the three is part of a larger virialized system, though there is
some indication that IC~1531 is in a region in which a poor group may form
in the future. The three galaxies have a range of X-ray properties. NGC~57
has an extended halo of $\sim$0.9~keV, roughly solar metallicity gas.
IC~1531 is a radio galaxy and its X-ray emission is dominated by the AGN
and an associated jet, with only a compact, cooler ($\sim$0.5~keV) halo.
NGC~7796 has no strong AGN, and possesses a halo with a similar temperature
to that of IC~1531, but which is more extended.

For each galaxy we model the surface brightness and temperature
distributions of the gas haloes. In IC~1531 the compactness of the halo and
the strength of the AGN contribution limits us to two radial temperature
bins, but three and four bins are used for NGC~57 and NGC~7796
respectively. All three galaxies are consistent with having isothermal
temperature profiles, though there is some evidence of a slight drop in the
central temperature of NGC~57. For NGC~7796 and NGC~57 we use these
measurements to estimate the gas content, cooling time and entropy of the
haloes and perform a comparison with NGC~4555, another elliptical detailed
in a previous paper.  The three galaxies have similar entropy and cooling
time profiles, but NGC~7796 has a slightly lower gas mass at all radii.

We also estimate the total mass and mass-to-light ratio profiles of NGC~57
and NGC~7796, and find that NGC~57 has very similar properties to NGC~4555;
both galaxies have fairly massive dark haloes. The M/L ratio of NGC~57 at
4.75$r_e$ is found to be 44.7$^{+4.0}_{-8.5}$~\ML\ (1$\sigma$
uncertainties). This demonstrates that NGC~4555 is not a unique system, and
strongly suggests that elliptical galaxies can possess moderately massive
dark haloes independent of larger-scale systems. At present NGC~4555 and
NGC~57 are the two most isolated ellipticals with accurate X-ray mass
profiles, and NGC~821 is the only comparably isolated galaxy with a
dynamical mass estimate. NGC~7796 has a lower total mass and M/L ratio at
any given radius, with a M/L of 10.6$^{+2.5}_{-2.3}$ \ML\ at 5$r_e$. This
may indicate that the dark matter halo of the galaxy is not as massive as
that of NGC~57, or the structure of their DM their profiles differ.
However, very low M/L ratios are found in the core of the galaxy, and this
may indicate that our X-ray-derived mass is unreliable because the gas is
not in hydrostatic equilibrium, perhaps because it takes the form of a
galactic wind.  Comparison with other X-ray and dynamical mass estimates
shows a number of systems with values comparable to those found for our
galaxies, but also reveals a number of factors which may have biased the
results.  Cross-comparison of mass estimates for the same object using
independent techniques seems the mostly likely way to resolve these issues.

\vspace{.5cm}
\noindent{\textbf{acknowledgments}}\\
We thank the referee, Y. Fukazawa, for comments which have materially
improved the paper.  Support for this work was provided by the National
Aeronautics and Space Administration through NASA Grants NNG04GF19G and
NNG05GI62G, and through Chandra Award Numbers G05-6129X and G06-7070X-R
issued by the Chandra X-ray Observatory Center, which is operated by the
Smithsonian Astrophysical Observatory for and on behalf of the National
Aeronautics Space Administration under contract NAS8-03060. AJRS also
acknowledges support from PPARC. This research made use of the NASA/IPAC
Extragalactic Database (NED) which is operated by the Jet Propulsion
Laboratory, California Institute of Technology, under contract with the
National Aeronautics and Space Administration, the Digitized Sky Surveys
(DSS), produced at the Space Telescope Science Institute under U.S.
Government grant NAG W-2166, NASA's Astrophysics Data System, and the
HyperLeda database (http://leda.univ-lyon1.fr).

\bibliographystyle{mn2e}
\bibliography{../../paper}

\begin{thebibliography}{}

\bibitem[\protect\citeauthoryear{{Acreman}, {Stevens}, {Ponman} \&
  {Sakelliou}}{{Acreman} et~al.}{2003}]{Acremanetal03}
{Acreman} D.~M.,  {Stevens} I.~R.,  {Ponman} T.~J.,    {Sakelliou} I.,  2003,
  MNRAS, 341, 1333

\bibitem[\protect\citeauthoryear{{Anders} \& {Grevesse}}{{Anders} \&
  {Grevesse}}{1989}]{AndersGrevesse89}
{Anders} E.,  {Grevesse} N.,  1989, Geo.~et~Cosmo.~Acta, 53, 197

\bibitem[\protect\citeauthoryear{{Arnaud}, {Majerowicz}, {Lumb}, {Neumann},
  {Aghanim}, {Blanchard}, {Boer}, {Burke}, {Collins}, {Giard}, {Nevalainen},
  {Nichol}, {Romer} \& {Sadat}}{{Arnaud} et~al.}{2002}]{Arnaudetal02}
{Arnaud} M.,  {Majerowicz} S.,  {Lumb} D.,  {Neumann} D.~M.,  {Aghanim} N.,
  {Blanchard} A.,  {Boer} M.,  {Burke} D.~J.,  {Collins} C.~A.,  {Giard} M.,
  {Nevalainen} J.,  {Nichol} R.~C.,  {Romer} A.~K.,    {Sadat} R.,  2002, A\&A,
  390, 27

\bibitem[\protect\citeauthoryear{{Athey}, {Bregman}, {Bregman}, {Temi} \&
  {Sauvage}}{{Athey} et~al.}{2002}]{Atheyetal02}
{Athey} A.,  {Bregman} J.,  {Bregman} J.,  {Temi} P.,    {Sauvage} M.,  2002,
  ApJ, 571, 272

\bibitem[\protect\citeauthoryear{{Bacon}, {Monnet} \& {Simien}}{{Bacon}
  et~al.}{1985}]{Baconetal85}
{Bacon} R.,  {Monnet} G.,    {Simien} F.,  1985, A\&A, 152, 315

\bibitem[\protect\citeauthoryear{{Bahcall} \& {Tremaine}}{{Bahcall} \&
  {Tremaine}}{1981}]{BahcallTremaine81}
{Bahcall} J.~N.,  {Tremaine} S.,  1981, ApJ, 244, 805

\bibitem[\protect\citeauthoryear{{Barnes}}{{Barnes}}{1989}]{Barnes89}
{Barnes} J.~E.,  1989, Nature, 338, 123

\bibitem[\protect\citeauthoryear{{Bertin}, {Bertola}, {Buson}, {Danzinger},
  {Dejonghe}, {Sadler}, {Saglia}, {de Zeeuw} \& {Zeilinger}}{{Bertin}
  et~al.}{1994}]{Bertinetal94}
{Bertin} G.,  {Bertola} F.,  {Buson} L.~M.,  {Danzinger} I.~J.,  {Dejonghe} H.,
   {Sadler} E.~M.,  {Saglia} R.~P.,  {de Zeeuw} P.~T.,    {Zeilinger} W.~W.,
  1994, A\&A, 292, 381

\bibitem[\protect\citeauthoryear{{B{\^i}rzan}, {Rafferty}, {McNamara}, {Wise}
  \& {Nulsen}}{{B{\^i}rzan} et~al.}{2004}]{Birzanetal04}
{B{\^i}rzan} L.,  {Rafferty} D.~A.,  {McNamara} B.~R.,  {Wise} M.~W.,
  {Nulsen} P.~E.~J.,  2004, ApJ, 607, 800

\bibitem[\protect\citeauthoryear{Buote \& Fabian}{Buote \&
  Fabian}{1998}]{Buotefabian98}
Buote D.,  Fabian A.,  1998, MNRAS, 296, 977

\bibitem[\protect\citeauthoryear{{Buote}}{{Buote}}{2000}]{Buote00b}
{Buote} D.~A.,  2000, MNRAS, 311, 176

\bibitem[\protect\citeauthoryear{Cash}{Cash}{1979}]{Cash79}
Cash W.,  1979, ApJ, 228, 939

\bibitem[\protect\citeauthoryear{Ciotti, D'Ercole, Pelegrini \& Renzini}{Ciotti
  et~al.}{1991}]{Ciottietal91}
Ciotti L.,  D'Ercole A.,  Pelegrini S.,    Renzini A.,  1991, ApJ, 376, 380

\bibitem[\protect\citeauthoryear{{Colless}, {Dalton}, {Maddox}, {Sutherland},
  {Norberg}, {Cole}, {Bland-Hawthorn}, {Bridges}, {Cannon}, {Collins}, {Couch},
  {Cross}, {Deeley} \& {et al.}}{{Colless} et~al.}{2001}]{Collessetal01}
{Colless} M.,  {Dalton} G.,  {Maddox} S.,  {Sutherland} W.,  {Norberg} P.,
  {Cole} S.,  {Bland-Hawthorn} J.,  {Bridges} T.,  {Cannon} R.,  {Collins} C.,
  {Couch} W.,  {Cross} N.,  {Deeley} K.,    {et al.} 2001, MNRAS, 328, 1039

\bibitem[\protect\citeauthoryear{{Dekel}, {Stoehr}, {Mamon}, {Cox} \&
  {Primack}}{{Dekel} et~al.}{2005}]{Dekeletal05}
{Dekel} A.,  {Stoehr} F.,  {Mamon} G.~A.,  {Cox} T.~J.,    {Primack} J.~R.,
  2005, Nature, 437, 707

\bibitem[\protect\citeauthoryear{{Diehl} \& {Statler}}{{Diehl} \&
  {Statler}}{2006}]{DiehlStatler07}
{Diehl} S.,  {Statler} T.~S.,  2006, ArXiv Astrophysics e-prints,
  astro-ph/0606215

\bibitem[\protect\citeauthoryear{{Douglas}, {Napolitano}, {Romanowsky} \&
  {Coccato}}{{Douglas} et~al.}{2007}]{Douglasetal07}
{Douglas} N.~G.,  {Napolitano} N.~R.,  {Romanowsky} A.~J.,    {Coccato} L.,
  2007, ApJ, submitted

\bibitem[\protect\citeauthoryear{Dressler}{Dressler}{1980}]{Dressler80}
Dressler A.,  1980, ApJ, 236, 351

\bibitem[\protect\citeauthoryear{{Ekers}, {Wall}, {Shaver}, {Goss}, {Fosbury},
  {Danziger}, {Moorwood}, {Malin}, {Monk} \& {Ekers}}{{Ekers}
  et~al.}{1989}]{Ekersetal89}
{Ekers} R.~D.,  {Wall} J.~V.,  {Shaver} P.~A.,  {Goss} W.~M.,  {Fosbury}
  R.~A.~E.,  {Danziger} I.~J.,  {Moorwood} A.~F.~M.,  {Malin} D.~F.,  {Monk}
  A.~S.,    {Ekers} J.~A.,  1989, MNRAS, 236, 737

\bibitem[\protect\citeauthoryear{Forman, Jones \& Tucker}{Forman
  et~al.}{1985}]{Formanjonestucker85}
Forman W.,  Jones C.,    Tucker W.,  1985, ApJ, 293, 102

\bibitem[\protect\citeauthoryear{{Freedman}, {Madore}, {Gibson}, {Ferrarese},
  {Kelson}, {Sakai}, {Mould}, {Kennicutt} Jr., {Ford}, {Graham}, {Huchra},
  {Hughes}, {Illingworth}, {Macri} \& {Stetson}}{{Freedman}
  et~al.}{2001}]{Freedmanetal01}
{Freedman} W.~L.,  {Madore} B.~F.,  {Gibson} B.~K.,  {Ferrarese} L.,  {Kelson}
  D.~D.,  {Sakai} S.,  {Mould} J.~R.,  {Kennicutt} Jr. R.~C.,  {Ford} H.~C.,
  {Graham} J.~A.,  {Huchra} J.~P.,  {Hughes} S.~M.~G.,  {Illingworth} G.~D.,
  {Macri} L.~M.,    {Stetson} P.~B.,  2001, ApJ, 553, 47

\bibitem[\protect\citeauthoryear{{Fruscione}, {McDowell}, {Allen},
  {Brickhouse}, {Burke}, {Davis}, {Durham}, {Elvis}, {Galle}, {Harris},
  {Huenemoerder}, {Houck} \& {et al.}}{{Fruscione}
  et~al.}{2006}]{Fruscioneetal06}
{Fruscione} A.,  {McDowell} J.~C.,  {Allen} G.~E.,  {Brickhouse} N.~S.,
  {Burke} D.~J.,  {Davis} J.~E.,  {Durham} N.,  {Elvis} M.,  {Galle} E.~C.,
  {Harris} D.~E.,  {Huenemoerder} D.~P.,  {Houck} J.~C.,    {et al.} 2006, in
  Observatory Operations: Strategies, Processes, and Systems. Edited by Silva,
  David R.; Doxsey, Rodger E.. Proceedings of the SPIE, Volume 6270, pp.
  (2006). {CIAO: Chandra's data analysis system}

\bibitem[\protect\citeauthoryear{{Fukazawa}, {Botoya-Nonesa}, {Pu}, {Ohto} \&
  {Kawano}}{{Fukazawa} et~al.}{2006}]{Fukazawaetal06}
{Fukazawa} Y.,  {Botoya-Nonesa} J.~G.,  {Pu} J.,  {Ohto} A.,    {Kawano} N.,
  2006, ApJ, 636, 698

\bibitem[\protect\citeauthoryear{Garcia}{Garcia}{1993}]{Garcia93}
Garcia A.~M.,  1993, A\&AS, 100, 47

\bibitem[\protect\citeauthoryear{{Gavazzi}, {Treu}, {Rhodes}, {Koopmans},
  {Bolton}, {Burles}, {Massey} \& {Moustakas}}{{Gavazzi}
  et~al.}{2007}]{Gavazzietal07}
{Gavazzi} R.,  {Treu} T.,  {Rhodes} J.~D.,  {Koopmans} L.~V.~E.,  {Bolton}
  A.~S.,  {Burles} S.,  {Massey} R.~J.,    {Moustakas} L.~A.,  2007, ArXiv
  Astrophysics e-prints, astro-ph/0701589

\bibitem[\protect\citeauthoryear{{Grevesse} \& {Sauval}}{{Grevesse} \&
  {Sauval}}{1998}]{GrevesseSauval98}
{Grevesse} N.,  {Sauval} A.~J.,  1998, Space Sci.~Rev., 85, 161

\bibitem[\protect\citeauthoryear{Helsdon, Ponman, O'Sullivan \& Forbes}{Helsdon
  et~al.}{2001}]{Helsdonetal01}
Helsdon S.~F.,  Ponman T.~J.,  O'Sullivan E.,    Forbes D.~A.,  2001, MNRAS,
  325, 693

\bibitem[\protect\citeauthoryear{{Humphrey}, {Buote}, {Gastaldello},
  {Zappacosta}, {Bullock}, {Brighenti} \& {Mathews}}{{Humphrey}
  et~al.}{2006}]{Humphreyetal06b}
{Humphrey} P.~J.,  {Buote} D.~A.,  {Gastaldello} F.,  {Zappacosta} L.,
  {Bullock} J.~S.,  {Brighenti} F.,    {Mathews} W.~G.,  2006, ApJ, 646, 899

\bibitem[\protect\citeauthoryear{{Jansen}, {Lumb}, {Altieri}, {Clavel}, {Ehle},
  {Erd}, {Gabriel}, {Guainazzi}, {Gondoin}, {Much}, {Munoz}, {Santos},
  {Schartel}, {Texier} \& {Vacanti}}{{Jansen} et~al.}{2001}]{Jansenetal01}
{Jansen} F.,  {Lumb} D.,  {Altieri} B.,  {Clavel} J.,  {Ehle} M.,  {Erd} C.,
  {Gabriel} C.,  {Guainazzi} M.,  {Gondoin} P.,  {Much} R.,  {Munoz} R.,
  {Santos} M.,  {Schartel} N.,  {Texier} D.,    {Vacanti} G.,  2001, A\&A, 365,
  L1

\bibitem[\protect\citeauthoryear{{Jensen}, {Tonry}, {Barris}, {Thompson},
  {Liu}, {Rieke}, {Ajhar} \& {Blakeslee}}{{Jensen} et~al.}{2003}]{Jensenetal03}
{Jensen} J.~B.,  {Tonry} J.~L.,  {Barris} B.~J.,  {Thompson} R.~I.,  {Liu}
  M.~C.,  {Rieke} M.~J.,  {Ajhar} E.~A.,    {Blakeslee} J.~P.,  2003, ApJ, 583,
  712

\bibitem[\protect\citeauthoryear{{Jones}, {Ponman}, {Horton}, {Babul},
  {Ebeling} \& {Burke}}{{Jones} et~al.}{2003}]{Jonesetal03}
{Jones} L.~R.,  {Ponman} T.~J.,  {Horton} A.,  {Babul} A.,  {Ebeling} H.,
  {Burke} D.~J.,  2003, MNRAS, 343, 627

\bibitem[\protect\citeauthoryear{Kellogg, Baldwin \& Koch}{Kellogg
  et~al.}{1975}]{Kelloggetal75}
Kellogg E.,  Baldwin J.~R.,    Koch D.,  1975, ApJ, 199, 299

\bibitem[\protect\citeauthoryear{{Khosroshahi}, {Jones} \&
  {Ponman}}{{Khosroshahi} et~al.}{2004}]{Khosroshahietal04}
{Khosroshahi} H.~G.,  {Jones} L.~R.,    {Ponman} T.~J.,  2004, MNRAS, 349, 1240

\bibitem[\protect\citeauthoryear{{Mamon} \& {{\L}okas}}{{Mamon} \&
  {{\L}okas}}{2005a}]{MamonLokas05a}
{Mamon} G.~A.,  {{\L}okas} E.~L.,  2005a, MNRAS, 362, 95

\bibitem[\protect\citeauthoryear{{Mamon} \& {{\L}okas}}{{Mamon} \&
  {{\L}okas}}{2005b}]{MamonLokas05b}
{Mamon} G.~A.,  {{\L}okas} E.~L.,  2005b, MNRAS, 363, 705

\bibitem[\protect\citeauthoryear{{Mathews} \& {Brighenti}}{{Mathews} \&
  {Brighenti}}{2003}]{MathewsBrighenti03}
{Mathews} W.~G.,  {Brighenti} F.,  2003, ARA\&A, 41, 191

\bibitem[\protect\citeauthoryear{{Melnick} \& {Sargent}}{{Melnick} \&
  {Sargent}}{1977}]{MelnickSargent77}
{Melnick} J.,  {Sargent} W.~L.~W.,  1977, ApJ, 215, 401

\bibitem[\protect\citeauthoryear{{Milone}, {Rickes} \& {Pastoriza}}{{Milone}
  et~al.}{2007}]{Miloneetal07}
{Milone} A.~d.~C.,  {Rickes} M.~G.,    {Pastoriza} M.~G.,  2007, A\&A, 469, 89

\bibitem[\protect\citeauthoryear{{Napolitano}, {Capaccioli}, {Romanowsky},
  {Douglas}, {Merrifield}, {Kuijken}, {Arnaboldi}, {Gerhard} \&
  {Freeman}}{{Napolitano} et~al.}{2005}]{Napolitanoetal05b}
{Napolitano} N.~R.,  {Capaccioli} M.,  {Romanowsky} A.~J.,  {Douglas} N.~G.,
  {Merrifield} M.~R.,  {Kuijken} K.,  {Arnaboldi} M.,  {Gerhard} O.,
  {Freeman} K.~C.,  2005, MNRAS, 357, 691

\bibitem[\protect\citeauthoryear{{O'Sullivan}, {Forbes} \&
  {Ponman}}{{O'Sullivan} et~al.}{2001}]{OFP01cat}
{O'Sullivan} E.,  {Forbes} D.~A.,    {Ponman} T.~J.,  2001, MNRAS, 328, 461

\bibitem[\protect\citeauthoryear{{O'Sullivan} \& {Ponman}}{{O'Sullivan} \&
  {Ponman}}{2004}]{OSullivanPonman04b_special}
{O'Sullivan} E.,  {Ponman} T.~J.,  2004, MNRAS, 354, 935, OP04

\bibitem[\protect\citeauthoryear{{O'Sullivan}, {Ponman} \&
  {Collins}}{{O'Sullivan} et~al.}{2003}]{OPC03}
{O'Sullivan} E.,  {Ponman} T.~J.,    {Collins} R.~S.,  2003, MNRAS, 340, 1375

\bibitem[\protect\citeauthoryear{{O'Sullivan}, {Vrtilek}, {Harris} \&
  {Ponman}}{{O'Sullivan} et~al.}{2007}]{OSullivanetal06}
{O'Sullivan} E.,  {Vrtilek} J.~M.,  {Harris} D.~E.,    {Ponman} T.~J.,  2007,
  ApJ, 658, 299

\bibitem[\protect\citeauthoryear{{O'Sullivan}, {Vrtilek}, {Kempner}, {David} \&
  {Houck}}{{O'Sullivan} et~al.}{2005}]{OSullivanetal05}
{O'Sullivan} E.,  {Vrtilek} J.~M.,  {Kempner} J.~C.,  {David} L.~P.,    {Houck}
  J.~C.,  2005, MNRAS, 357, 1134

\bibitem[\protect\citeauthoryear{{Paturel}, {Petit}, {Prugniel}, {Theureau},
  {Rousseau}, {Brouty}, {Dubois} \& {Cambr{\'e}sy}}{{Paturel}
  et~al.}{2003}]{Patureletal03}
{Paturel} G.,  {Petit} C.,  {Prugniel} P.,  {Theureau} G.,  {Rousseau} J.,
  {Brouty} M.,  {Dubois} P.,    {Cambr{\'e}sy} L.,  2003, A\&A, 412, 45

\bibitem[\protect\citeauthoryear{{Pellegrini}, {Baldi}, {Kim}, {Fabbiano},
  {Soria}, {Siemiginowska} \& {Elvis}}{{Pellegrini}
  et~al.}{2007}]{Pellegrinietal07}
{Pellegrini} S.,  {Baldi} A.,  {Kim} D.~W.,  {Fabbiano} G.,  {Soria} R.,
  {Siemiginowska} A.,    {Elvis} M.,  2007, ArXiv Astrophysics e-prints,
  astro-ph/0701639

\bibitem[\protect\citeauthoryear{Pellegrini \& Ciotti}{Pellegrini \&
  Ciotti}{1998}]{Pellegriniciotti98}
Pellegrini S.,  Ciotti L.,  1998, A\&A, 333, 433

\bibitem[\protect\citeauthoryear{{Pellegrini} \& {Ciotti}}{{Pellegrini} \&
  {Ciotti}}{2006}]{PellegriniCiotti06}
{Pellegrini} S.,  {Ciotti} L.,  2006, MNRAS, 370, 1797

\bibitem[\protect\citeauthoryear{Ponman \& Bertram}{Ponman \&
  Bertram}{1993}]{Ponmanbertram93}
Ponman T.~J.,  Bertram D.,  1993, Nature, 363, 51

\bibitem[\protect\citeauthoryear{{Pratt}, {Arnaud} \& {Aghanim}}{{Pratt}
  et~al.}{2001}]{Prattetal01}
{Pratt} G.~W.,  {Arnaud} M.,    {Aghanim} N.,  2001, in {Neumann} D.~M.,
  {Tranh Thanh Van} J.,  eds, Clusters of Galaxies and the High Redshift
  Universe Observed in X-rays: {XMM-Newton observations of galaxy clusters; the
  radial temperature profile of A2163}

\bibitem[\protect\citeauthoryear{{Reda}, {Forbes} \& {Hau}}{{Reda}
  et~al.}{2005}]{Redaetal05}
{Reda} F.~M.,  {Forbes} D.,    {Hau} G.~T.~K.,  2005, MNRAS, 360, 693

\bibitem[\protect\citeauthoryear{{Reda}, {Forbes}, {Beasley}, {O'Sullivan} \&
  {Goudfrooij}}{{Reda} et~al.}{2004}]{Redaetal04}
{Reda} F.~M.,  {Forbes} D.~A.,  {Beasley} M.~A.,  {O'Sullivan} E.~J.,
  {Goudfrooij} P.,  2004, MNRAS, 354, 851

\bibitem[\protect\citeauthoryear{{Reiprich} \& {B{\" o}hringer}}{{Reiprich} \&
  {B{\" o}hringer}}{2002}]{ReiprichBohringer02}
{Reiprich} T.~H.,  {B{\" o}hringer} H.,  2002, ApJ, 567, 716

\bibitem[\protect\citeauthoryear{{Romanowsky}, {Douglas}, {Arnaboldi},
  {Kuijken}, {Merrifield}, {Napolitano}, {Capaccioli} \&
  {Freeman}}{{Romanowsky} et~al.}{2003}]{Romanowskyetal03}
{Romanowsky} A.~J.,  {Douglas} N.~G.,  {Arnaboldi} M.,  {Kuijken} K.,
  {Merrifield} M.~R.,  {Napolitano} N.~R.,  {Capaccioli} M.,    {Freeman}
  K.~C.,  2003, Science, 301, 1696

\bibitem[\protect\citeauthoryear{Sanderson, Ponman, Finoguenov, Lloyd-Davies \&
  Markevitch}{Sanderson et~al.}{2003}]{Sandersonetal03}
Sanderson A. J.~R.,  Ponman T.~J.,  Finoguenov A.,  Lloyd-Davies E.~J.,
  Markevitch M.,  2003, MNRAS, 340, 989

\bibitem[\protect\citeauthoryear{{Smith}, {Brickhouse}, {Liedahl} \&
  {Raymond}}{{Smith} et~al.}{2001}]{Smithetal01}
{Smith} R.~K.,  {Brickhouse} N.~S.,  {Liedahl} D.~A.,    {Raymond} J.~C.,
  2001, ApJ, 556, L91

\bibitem[\protect\citeauthoryear{{Smith}, {Mart{\'{\i}}nez} \&
  {Graham}}{{Smith} et~al.}{2004}]{Smithetal04}
{Smith} R.~M.,  {Mart{\'{\i}}nez} V.~J.,    {Graham} M.~J.,  2004, ApJ, 617,
  1017

\bibitem[\protect\citeauthoryear{{Spergel}, {Bean}, {Dore'}, {Nolta},
  {Bennett}, {Hinshaw}, {Jarosik}, {Komatsu}, {Page}, {Peiris}, {Verde},
  {Barnes}, {Halpern} \& {et al.}}{{Spergel} et~al.}{2007}]{Spergeletal07}
{Spergel} D.~N.,  {Bean} R.,  {Dore'} O.,  {Nolta} M.~R.,  {Bennett} C.~L.,
  {Hinshaw} G.,  {Jarosik} N.,  {Komatsu} E.,  {Page} L.,  {Peiris} H.~V.,
  {Verde} L.,  {Barnes} C.,  {Halpern} M.,    {et al.} 2007, ApJS, 170, 377

\bibitem[\protect\citeauthoryear{{Sun}, {Jerius} \& {Jones}}{{Sun}
  et~al.}{2005}]{Sunetal05b}
{Sun} M.,  {Jerius} D.,    {Jones} C.,  2005, ApJ, 633, 165

\bibitem[\protect\citeauthoryear{{Sun}, {Vikhlinin}, {Forman}, {Jones} \&
  {Murray}}{{Sun} et~al.}{2005}]{Sunetal05a}
{Sun} M.,  {Vikhlinin} A.,  {Forman} W.,  {Jones} C.,    {Murray} S.~S.,  2005,
  ApJ, 619, 169

\bibitem[\protect\citeauthoryear{{Thomas}, {Maraston}, {Bender} \& {Mendes de
  Oliveira}}{{Thomas} et~al.}{2005}]{Thomasetal05}
{Thomas} D.,  {Maraston} C.,  {Bender} R.,    {Mendes de Oliveira} C.,  2005,
  ApJ, 621, 673

\bibitem[\protect\citeauthoryear{{Tonry}, {Dressler}, {Blakeslee}, {Ajhar},
  {Fletcher}, {Luppino}, {Metzger} \& {Moore}}{{Tonry}
  et~al.}{2001}]{Tonryetal01}
{Tonry} J.~L.,  {Dressler} A.,  {Blakeslee} J.~P.,  {Ajhar} E.~A.,  {Fletcher}
  A.~B.,  {Luppino} G.~A.,  {Metzger} M.~R.,    {Moore} C.~B.,  2001, ApJ, 546,
  681

\bibitem[\protect\citeauthoryear{{Trinchieri}, {Fabbiano} \&
  {Canizares}}{{Trinchieri} et~al.}{1986}]{Trinchierietal86}
{Trinchieri} G.,  {Fabbiano} G.,    {Canizares} C.~R.,  1986, ApJ, 310, 637

\bibitem[\protect\citeauthoryear{{Tully}}{{Tully}}{1987}]{Tully87}
{Tully} R.~B.,  1987, ApJ, 321, 280

\bibitem[\protect\citeauthoryear{{Voit} \& {Donahue}}{{Voit} \&
  {Donahue}}{2005}]{VoitDonahue05}
{Voit} G.~M.,  {Donahue} M.,  2005, ApJ, 634, 955

\bibitem[\protect\citeauthoryear{{Wegner}, {Bernardi}, {Willmer}, {da Costa},
  {Alonso}, {Pellegrini}, {Maia}, {Chaves} \& {Rit{\'e}}}{{Wegner}
  et~al.}{2003}]{Wegneretal03}
{Wegner} G.,  {Bernardi} M.,  {Willmer} C.~N.~A.,  {da Costa} L.~N.,  {Alonso}
  M.~V.,  {Pellegrini} P.~S.,  {Maia} M.~A.~G.,  {Chaves} O.~L.,    {Rit{\'e}}
  C.,  2003, AJ, 126, 2268

\bibitem[\protect\citeauthoryear{{Weisskopf}, {Brinkman}, {Canizares},
  {Garmire}, {Murray} \& {Van Speybroeck}}{{Weisskopf}
  et~al.}{2002}]{Weisskopfetal02}
{Weisskopf} M.~C.,  {Brinkman} B.,  {Canizares} C.,  {Garmire} G.,  {Murray}
  S.,    {Van Speybroeck} L.~P.,  2002, PASP, 114, 1

\bibitem[\protect\citeauthoryear{{Zezas}, {Birkinshaw}, {Worrall}, {Peters} \&
  {Fabbiano}}{{Zezas} et~al.}{2005}]{Zezasetal05}
{Zezas} A.,  {Birkinshaw} M.,  {Worrall} D.~M.,  {Peters} A.,    {Fabbiano} G.,
   2005, ApJ, 627, 711

\end{thebibliography}

\label{lastpage}
\end{document}